\documentclass{article}

\usepackage[a4paper,top=3cm,bottom=3cm,left=3cm,right=3cm]{geometry}

\usepackage{authblk}
\usepackage{amsthm,amsmath,amsfonts,amssymb}
\usepackage[colorlinks,allcolors=blue]{hyperref}
\usepackage[ruled,vlined]{algorithm2e}
\usepackage{bbm}
\usepackage{thmtools}
\usepackage[nameinlink]{cleveref}
\usepackage[normalem]{ulem}
\usepackage{xparse}
\usepackage{tabularx,booktabs}
\usepackage{multirow}
\usepackage[section]{placeins} 

\usepackage[normalem]{ulem}
\usepackage[doublespacing]{setspace}
\usepackage{bm}
\usepackage{graphicx}
\usepackage{pgfmath}
\usepackage{subfig}
\usepackage{siunitx}
\usepackage{xcolor}

\definecolor{darkgreen}{rgb}{0,0.8,0} \definecolor{darkred}{rgb}{0.8,0,0}

\newcommand{\ScaleMax}{0.50}

\DeclareMathOperator{\E}{E}

\DeclareMathOperator{\Var}{Var}

\newcommand{\numprint}[1]{\num{#1}}
\newcommand{\coef}[2]{
 \pgfmathsetmacro{\absval}{abs(#1)}
 \pgfmathsetmacro{\rawperc}{100*(\absval)/(\ScaleMax)}
 \pgfmathsetmacro{\perc}{min(100, round(\rawperc))}
 \ifnum#2=0
 \textcolor{black}{\numprint{#1}}
 \else
 \ifdim #1pt > 0pt
 \textcolor{green!\perc!black}{\numprint{#1}}
 \else
 \textcolor{red!\perc!black}{\numprint{#1}}
 \fi
 \fi
}

\newcommand{\virgo}[1]{``#1''}
\newcommand{\iid}{\stackrel{\rm iid}{\sim}}
\newcommand{\ind}{\stackrel{\rm ind}{\sim}}

\usepackage{hyperref}

\usepackage{fontawesome5} 
\usepackage{xcolor}
\usepackage[authoryear]{natbib}\definecolor{orcidgreen}{HTML}{A6CE39}

\newcommand{\ORCID}[1]{\href{https://orcid.org/#1}{\textcolor{orcidgreen}{\faOrcid}}}

\def\keywords#1{%
\if@modern%
    \vspace*{6pt}%
\else%
\if@traditional
    \vspace*{12pt}
\else
    \vspace*{8pt}
\fi
\fi\par%
\if@modern
{\sffamilyfontcnbold\fontsize{8bp}{11}\keywordsname\ \sffamilyfontcn#1}%
\else%
\if@traditional
{\fontsize{8bp}{10}\textbf{\uppercase{Keywords}:}\ #1}%
\else
{\fontsize{8bp}{10}\textbf{{Key words}:}\ #1}%
\fi\fi
}%
    

\title{\textbf{A Bayesian multivariate longitudinal framework for preventive screening of metabolic syndrome in blood donors}}
\author[1]{Simone Colombara\protect\ORCID{0009-0007-4522-7988
}}
\author[1]{Ilenia Epifani\protect\ORCID{0000-0001-9005-1378}}
\author[1]{Alessandra Guglielmi\protect\ORCID{0000-0001-7005-7588}}
\author[2]{Ettore Lanzarone\protect\ORCID{0000-0001-8816-9086}}
\affil[1]{\small\textit{Department of Mathematics, Politecnico di Milano, Milan, Italy}}
\affil[2]{\small\textit{Department of Management, Information and Production Engineering, University of Bergamo, Dalmine (BG), Italy}}

\date{} 

\begin{document}

\maketitle


\begin{abstract}
Metabolic syndrome (MetS) is a major public health concern characterised by the co-occurrence of multiple metabolic risk factors that may remain undetected for long periods.
Longitudinal monitoring of predominantly healthy individuals may facilitate the early identification of elevated metabolic risk.
We propose a Bayesian statistical framework to estimate the probability that a blood donor will meet the diagnostic criteria for MetS at a current or future pre-donation assessment, using information collected at previous visits and covariates available at the time of assessment.
Using longitudinal data from AVIS Milan, one of the main blood donor associations in Italy, we analyse repeated clinical and lifestyle measurements from a predominantly healthy donor population.
We fit a Bayesian multivariate mixed-effects model that jointly represents the logarithms of the five diagnostic components of MetS, accounting for dependence among components and repeated measurements within donors.
The posterior predictive probabilities provide individualised risk estimates, which are translated into an interpretable traffic-light warning system (low, intermediate, and high risk) to support clinical prioritisation during pre-donation screening.
The framework also allows clinicians to explore hypothetical scenarios involving modifiable risk factors and assess how predicted MetS risk changes under alternative covariate profiles, providing a potential tool for personalised preventive counselling.
\end{abstract}
\keywords{Bayesian multivariate mixed-effects models, longitudinal data, low-prevalence population, onset prediction, posterior predictive inference, 
traffic-light classification}

\section{Introduction}
\label{sec:introduction}

Metabolic syndrome (MetS) is a complex clinical condition characterised by the simultaneous presence of multiple metabolic abnormalities, including abdominal obesity, elevated blood pressure, high fasting glucose, elevated triglycerides, and low levels of HDL cholesterol.
These interrelated abnormalities, often associated with insulin resistance, jointly increase the risk of cardiovascular disease, type 2 diabetes, stroke, and mortality \citep{isomaa2001cardiovascular,najarian2006metabolic,air2007diabetes}.
\cite{noubiap2022geographic} estimate an increase of MetS prevalence of 3.4\% every ten years, reaching over 31.5\% prevalence in 2020s in the WHO European region.
Beyond its clinical severity, this trend imposes an economic burden on public health expenditures driven by MetS-associated conditions such as obesity and diabetes.
Globally, direct health expenditure for diabetes surpassed 1 trillion US\$ in 2024 \citep{idf_atlas_2025}, while the economic impact of obesity is projected to exceed 4 trillion US\$ annually by 2035 \citep{worldobesityatlas2023}.
To mitigate this load on health and public expenditure, developing data-driven tools for early screening and prevention of MetS is an urgent public health priority.

There are several definitions of MetS provided by different organisations, e.g., the International Diabetes Federation (IDF) and the National Cholesterol Education Panel-Adult Treatment Panel III (NCEP-ATP III) \citep{grundy2005diagnosis,alberti2006metabolic,kassi2011metabolic}.
While they share a common conceptual framework, they differ in diagnostic thresholds and sometimes include ethnicity-specific criteria.
In this study, we adopt the NCEP-ATP III definition of MetS, according to which a diagnosis is established when at least three of the conditions in Table~\ref{tab:MetS_criteria} are met.

\begin{table}[htbp]
\centering
\begin{tabular}{lcc}
\toprule
\textbf{Component} & \textbf{Men} & \textbf{Women} \\
\midrule
Waist circumference ($cm$) & $> 102$ & $> 88$ \\
Systolic Blood pressure ($mmHg$) & $\geq 130$ & $\geq 130$ \\
Fasting Glucose ($mg/dL$) & $\geq 100$ & $\geq 100$ \\
Triglycerides ($mg/dL$) & $\geq 150$ & $\geq 150$ \\
HDL cholesterol ($mg/dL$) & $< 40$ & $< 50$ \\
\bottomrule
\end{tabular}
\caption{MetS components according to the NCEP-ATP III definition, and diagnostic thresholds for men and women. The diagnosis of MetS is established when at least three conditions are met.}
\label{tab:MetS_criteria}
\end{table}

MetS often develops silently, with few or no symptoms in its early stages, and is rarely the primary focus of clinical screening.
Consequently, MetS and its individual components can remain undiagnosed for long periods, delaying potentially beneficial lifestyle interventions such as dietary changes, increased physical activity, and weight management \citep{grundy2005diagnosis}.
However, data-driven early diagnosis of MetS remains challenging, as most available health data pertain to individuals who have already developed the disease.
In contrast, data on healthy individuals before disease onset are scarce and accessible only in a few specific facilities.

Blood donation centers are one such setting, as donors undergo repeated blood tests over several years.
Typical eligibility criteria for blood donation exclude individuals with severe cardiovascular or systemic diseases; therefore, blood donors constitute a selected, generally healthy adult cohort.
Accordingly, blood donors have also been used as a reference group in comparative studies between healthy and clinical populations \citep{roer2025metabolic}.
Nonetheless, evidence from cross-sectional studies shows that MetS components, including abdominal obesity, dyslipidemia, impaired fasting glucose, and elevated blood pressure, are also present among blood donors \citep{peffer2015donation}.
Potential cardiovascular and metabolic effects of blood donation itself, particularly with respect to iron metabolism, have also been investigated, but the evidence remains inconclusive or contradictory \citep{salonen1998donation,ascherio2001blood,germain2013iron,peffer2015donation}.
A recent systematic review found no strong evidence for a clear association between frequent blood donation and a reduced risk of MetS-related diseases \citep{quee2022cardiovascular}.
Overall, blood donation centers provide a valuable source of repeated metabolic measurements in a healthy population.

\subsection{Our contribution}

We consider a large longitudinal dataset from one of the main blood donor associations in Italy, namely the Milan department of the \textit{Associazione Volontari Italiani del Sangue} (AVIS Milan), which contains repeated clinical measurements and lifestyle data.
On the day of donation, each donor first undergoes a pre-donation check-up with an AVIS Milan clinician, followed by blood donation and sample collection for laboratory testing.
This check-up represents the only direct consultation between the donor and the clinician during the donation visit and, importantly, takes place before the current laboratory results become available.
The blood samples are collected during the donation, while the corresponding laboratory results become available a few days later, once testing is complete, and are then communicated to the donor and recorded in the AVIS Milan database.
Thus, during the pre-donation consultation, the clinician has access to the five target variables and the other biomarkers measured at previous visits, including the most recent one, but not to their values at the current visit.
Because consecutive donations may be separated by several months, the latest available laboratory measurements may refer to a substantially earlier time point.

This paper proposes a Bayesian statistical framework to estimate the probability that a blood donor will meet the diagnostic criteria for MetS at the current (or a next) pre-donation assessment, leveraging information collected at previous visits and covariates available at the current (or a next) visit.
Our contribution is twofold.
First, we develop a Bayesian multivariate longitudinal model that jointly represents the logarithms of the five diagnostic components of MetS listed in Table~\ref{tab:MetS_criteria}.
The model accounts for both the dependence among these components and the within-donor dependence induced by repeated measurements.
In particular, we provide individualised probabilistic risk estimates, i.e.,
posterior predictive probabilities that a donor will meet the MetS criteria at a future time.
Second, to facilitate clinical interpretation and practical use, we translate the posterior predictive probabilities into an interpretable \virgo{traffic-light} risk classification.
Each donor is classified as having a low (green), intermediate (yellow), or high (red) risk of meeting the diagnostic criteria for MetS at the current or a future assessment.

Beyond providing individualised risk estimates, the framework allows clinicians to explore hypothetical scenarios that involve modifiable risk factors
and assess how the predicted MetS risk would vary under alternative covariate profiles.
The resulting tool may support preventive counselling during routine pre-donation check-ups {or at any other future time}
and help donors understand how modifiable risk factors are associated with their predicted future risk of MetS.
By focusing clinical attention on at-risk donors 
during  pre-donation consultations, 
our framework may support earlier preventive interventions without substantially increasing screening costs.
Nevertheless, the framework is not limited to blood donation and may apply to screening predominantly healthy individuals in other contexts, provided longitudinal measurements are available.

The Bayesian model adopted in this paper results from a comprehensive model-building strategy aimed at appropriately representing the multivariate and longitudinal dependence among the five MetS components, while also accounting for the temporal evolution of covariates.
Specifically, we progressively evaluated whether a multivariate formulation is preferable to independent univariate models, whether the current metabolic profile retains residual dependence on measurements from the previous donation, and 
whether the effects of time-varying covariates should vary with the inter-donation interval.
Overall, the multivariate mixed-effects model described in Section~\ref{sec:mult_GLMM} achieved the best balance between predictive accuracy, parsimony, computational efficiency, and clinical interpretability.

\subsection{Related methodological work}

To our knowledge, research on MetS among blood donors is largely cross-sectional and descriptive, with few studies leveraging large cohorts observed repeatedly over extended follow-up periods.
From a methodological perspective, longitudinal studies of MetS require models capable of jointly representing multiple correlated biomarkers measured over time.
The diagnostic components of MetS (see Table~\ref{tab:MetS_criteria}) are biologically interconnected and driven by common pathophysiological processes.
Analysing these components using separate univariate models does not explicitly represent their dependence and may therefore fail to fully exploit their joint dynamics \citep{rizopoulos2014combining}; see also \cite{RegitzZagrosek2016GenderCVD} for related considerations in the cardiovascular literature.
Longitudinal modelling strategies based on repeated biomarker measurements have also been investigated for early disease detection, for example in women at risk of ovarian cancer, although mainly in a univariate setting \citep{han2020statistical}.
In parallel, machine learning (ML) approaches, such as decision trees, random forests, gradient boosting, and deep learning architectures, have been widely applied to the prediction and classification of MetS, typically treating MetS diagnosis as a supervised classification problem based on cross-sectional data \citep[e.g.,][]{Park2021MetSMLSC, kim2022MetSMLBMC,diagnostics12123117,Shin2023MLMetS}.
However, these studies do not always focus on blood donors or other predominantly healthy cohorts, nor do they always incorporate invasive biomarkers.

A notable development is the multitask deep learning model proposed by \cite{lee2024MultiTaskMetS}, which jointly predicts MetS and its components using data from more than 70,000 adult participants in the Korean Genome and Epidemiology Study (KoGES) and outperforms several standard classifiers.
However, the model operates on dichotomised outcomes, does not explicitly represent the longitudinal dependence among repeated biomarker measurements, and provides limited quantification of predictive uncertainty.

Few ML studies incorporate longitudinal information.
Existing approaches, such as the visit-pair framework proposed by \cite{tavares2022prediction}, use measurements from one visit to predict MetS at the subsequent visit but do not model the joint longitudinal evolution of its components.
Other recent ML studies identify risk factors associated with MetS onset by selecting cohorts of initially healthy individuals and predicting their binary clinical status at the final visit \citep{yang2022machine,goldman2025predicting}.

Overall, existing ML approaches primarily target classification accuracy and generally do not jointly model the longitudinal trajectories of the MetS components or fully quantify predictive uncertainty.
These gaps motivate our Bayesian multivariate longitudinal approach to individualised risk estimation and early screening.
Bayesian generalized linear mixed models (GLMMs) offer a flexible and well-established framework for jointly modelling multiple longitudinal biomarkers.
Following the seminal work of \citet{ZegerKarim1991GLMMGibbs}, Bayesian GLMMs were extended to multivariate and longitudinal settings \citep[e.g.,][]{XuZeger2001Surrogate,Dunson2003DynamicLatent,cai2006bayesian}.

However, despite these methodological developments, such models have rarely been applied to MetS, whose diagnostic components have an intrinsically multivariate and longitudinal structure.
Related approaches were proposed by \citet{tayob2018multivariate,tayob2022multivariate} for hepatocellular carcinoma screening.
\cite{tayob2018multivariate} developed a fully Bayesian screening approach that combines the longitudinal trajectories of multiple non-invasive biomarkers to estimate an individual's posterior probability of disease and uses this probability to obtain a binary screening classification.
\cite{tayob2022multivariate} proposed a multivariate parametric empirical Bayes approach that uses an individual's longitudinal screening history to derive patient- and visit-specific biomarker thresholds while controlling the population-level false-positive rate.
Although these approaches were developed for an at-risk clinical population and a different disease, these studies illustrate the value of combining multivariate longitudinal biomarker information within probabilistic screening
frameworks.

\subsection{Organisation of the paper}

The paper is structured as follows.
Section~\ref{sec:data} introduces the dataset. 
Section~\ref{sec:mult_GLMM} describes the Bayesian multivariate mixed-effects model and discusses alternative models that account for the longitudinal evolution of both outcomes and covariates.
Posterior inference is performed using an MCMC algorithm implemented in Stan \citep{carpenter2017stan}, as detailed in Section~\ref{sec:results}.
Then, Section~\ref{sec:comparison} presents comparisons with alternative ML approaches and analyses the robustness of our model through a simulation study.
Section~\ref{sec:colloquium} describes the traffic-light warning system and illustrates how clinicians at AVIS Milan could communicate to donors their predicted risk of meeting the diagnostic criteria for MetS.
Finally, Section~\ref{sec:conclusion} concludes the paper by discussing the implications of our findings and the limitations of the study.

In addition, three appendices are included in a Supplementary Material file: Appendix~\ref{sec:appendixA} presents the preprocessing of the data and its exploratory analysis; Appendix~\ref{sec:appendixB} reports the alternative model specifications and time dependence structures considered for model assessment; Appendix~\ref{sec:appendixC} evaluates the performance of the proposed model on a simulated dataset.

\section{Data}
\label{sec:data}

We consider data from AVIS Milan, one of the largest voluntary blood donor organisations in Italy.
Blood donors at AVIS Milan must meet strict eligibility criteria established by Italian law, including minimum age and body-weight requirements, as well as acceptable ranges for physiological parameters such as blood pressure and haemoglobin concentration \citep{epifani2025predicting}.
The dataset integrates clinical records with self-reported lifestyle information.
Data sources include the EMONET medical database (mandatory for the Regional healthcare system), which contains information on donations and clinical exams for each donor, and the 
AVIS Milan internal database, which contains information on donors' lifestyles.
We select donors who made at least four donations between January 2019 and September 2023 to ensure sufficient data for parameter estimation in our longitudinal model.

The five target variables shown in Table~\ref{tab:MetS_criteria}, excluding \textit{waist circumference}, had missing rates below 1\%, and donors with missing values in these variables were excluded.
In this way, the dataset in this study comprises 11,556 visits across $I=2,228$ donors (1,907 men and 321 women).
As for \textit{waist circumference}, measurement is missing in 48.5\% of the visits, but these values are included as latent parameters in the Bayesian model described in Section~\ref{sec:mult_GLMM}.
As a consequence, at each iteration of the MCMC algorithm, we sample the missing values, together with the model parameters, by drawing directly from their full conditional distributions, conditional on the observed data and the current values of the other parameters.
This unified Bayesian formulation propagates missing-data uncertainty through posterior inference and avoids a two-stage procedure in which the same data
are used first for imputation and then for parameter estimation.

According to Italian regulations, the minimum interval between two consecutive donations is 90 days for men and post-menopausal women and 180 days for other women, with limited flexibility after clinical evaluation.
In this cohort, the average interval between consecutive visits is 291 days for men and 339 days for women.
These longer time gaps partly reflect the analysis of a time window that overlaps with the COVID-19 pandemic, which partially disrupted donation schedules. It is worth emphasising that the original NCEP-ATP III definition of MetS also included a threshold for diastolic pressure.
However, in this paper, we consider only \text{systolic blood pressure (\textit{PMAX})} rather than \text{diastolic blood pressure}, since the two are highly correlated and including both would result in redundant information.
This modification was discussed with AVIS Milan clinicians, who deemed it consistent with MetS characteristics and with the study's objective.

To implement a cross-validation approach, we divided the data into training, validation, and test sets.
The test set includes independent donors, i.e., donors not included in the other sets.
Specifically, we randomly selected 51 of the 2,228 donors using stratified sampling to preserve the original male-to-female ratio.
The remaining 2,177 donors were used to construct the training set and the validation set.
For each of these donors, the last available donation was reserved for the validation set, while all preceding donations were included in the training set, yielding 9,125 training observations.
Among these 2,177 donors, only 689 had complete information on \textit{waist circumference} at their last donation.

Exploratory data analysis on the training set is summarised below.
The initial dataset comprised 39 variables, including the five MetS targets.
We first removed covariates with high collinearity or missingness exceeding 50\% (see Supplementary Figure~\ref{fig:missing_values}).
Associations between covariates were assessed using chi-squared tests for categorical variables and empirical Pearson correlations for numerical variables.
Based on this exploratory analysis, we removed the variables \textit{haematocrit}, mean corpuscular volume (\textit{MCV}), and \textit{weight} to mitigate multicollinearity.
Among categorical variables, \textit{smoking} and \textit{alcohol} were discarded due to high missing rates and strong associations with other categorical predictors.
See Supplementary Appendix~\ref{app:eda} for additional details.
The blood group variables \textit{ABO} and \textit{Rh}, which showed limited predictive value for the metabolic targets (see Supplementary Table~\ref{tab:kw_target_cov}), were also excluded from the dataset.
We selected interaction terms using boxplots and Kruskal-Wallis tests between numerical and categorical covariates.
The final multivariate longitudinal model included 34 covariates and interaction terms, as reported in Table~\ref{tab:covariates-complete} along with their units of measure.
See Supplementary Appendix~\ref{app:eda} for details; in particular, Supplementary Figure~\ref{fig:hist_donation_count} shows the barplot of the donation counts for the training set.
\begin{table}[htbp]
\centering
\scriptsize
\begin{tabular}{llr}
\toprule
\textbf{Covariates} & \textbf{Empirical Mean $\pm$ SD} & \textbf{NA (\%)} \\
\midrule 
\textbf{Demographics} && \\
Age & 47.9 $\pm$ 10.00 \textit{years} & 0.0 \\
\textit{sex} & Men 86\% (1864 donors); Women 14\% (313 donors) & 0.0 \\
Height & 175.6 $\pm$ 7.68 \textit{cm} & 0.2 \\
\textit{physical activity} (Active/Inactive) & Active: 67\%; Inactive: 33\% & 43.9 \\
BMI & 25.4 $\pm$ 3.28 $\frac{kg}{m^2}$ & 0.2 \\
\midrule[-2pt]
\textbf{Measured by clinician} && \\
Heart Rate & 67.1 $\pm$ 7.07 \textit{bpm} & 0.1 \\
\midrule[-2pt]
\textbf{Blood tests} && \\
Hemoglobin & 14.9 $\pm$ 0.97 $\frac{g}{dL}$ & 0.1 \\[3pt]
Total Proteins & 7.1 $\pm$ 0.40 $\frac{g}{dL}$ & 30.2 \\[3pt]
Creatinine & 1.0 $\pm$ 0.15 $\frac{mg}{dL}$ & 28.9 \\[2pt]
Ferritin & 65.5 $\pm$ 49.14 $\frac{ng}{mL}$ & 22.1 \\[2pt]
Platelets & 236.0 $\pm$ 47.78 $\frac{10^3}{\mu L}$ & 0.1 \\[3pt]
Lymphocytes & 32.9 $\pm$ 6.87 (\%) & 0.1 \\
Eosinophils & 2.8 $\pm$ 1.97 (\%) & 0.1 \\
Basophils & 0.8 $\pm$ 0.32 (\%) & 0.1 \\
Monocytes & 9.0 $\pm$ 1.82 (\%) & 0.1 \\[2pt]
MCH & 29.3 $\pm$ 1.62 $\frac{pg}{cell}$ & 0.1 \\[2pt]
MCHC & 32.7 $\pm$ 0.98 $\frac{g}{d L}$ & 0.1 \\[2pt]
WBC & 5.9 $\pm$ 1.30 $\frac{10^3}{\mu L}$ & 0.1 \\
Blood Volume Distribution (\%) & 13.1 $\pm$ 0.89 & 0.1 \\
ALT & 21.3 $\pm$ 11.36 $\frac{units}{L}$ & 29.5 \\
\midrule[-2pt]
\textbf{Interactions} && \\
ALT $\times$ \textit{sex} & Men: 22.08 $\pm$ 11.41 $\frac{units}{L}$; Women 16.70 $\pm$ 9.78 $\frac{units}{L}$ & 29.5 \\[2pt]
Age $\times$ \textit{sex} & Men: 47.58 $\pm$ 9.76 \textit{years}; Women 50.21 $\pm$ 11.26 \textit{years} & 0.0 \\
BMI $\times$ \textit{sex} & Men: 25.56 $\pm$ 3.15 $\frac{kg}{m^2}$; Women 23.93 $\pm$ 3.75 $\frac{kg}{m^2}$ & 0.2 \\[4pt]
Creatinine $\times$ \textit{sex} & Men: 1.03 $\pm$ 0.14 $\frac{mg}{dL}$; Women 0.81 $\pm$ 0.12 $\frac{mg}{dL}$ & 28.9 \\[2pt]
Ferritin $\times$ \textit{sex} & Men: 68.04 $\pm$ 50.55 $\frac{ng}{mL}$; Women 48.80 $\pm$ 34.15 $\frac{ng}{mL}$ & 22.1 \\[2pt]
Height $\times$ \textit{sex} & Men: 177.39 $\pm$ 6.47 $cm$; Women 164.57 $\pm$ 5.84 $cm$ & 0.2 \\[2pt]
Heart Rate $\times$ \textit{sex} & Men: 66.93 $\pm$ 7.09 \textit{bpm}; Women 68.07 $\pm$ 6.90 \textit{bpm} & 0.1 \\[2pt]
Hemoglobin $\times$ \textit{sex} & Men: 15.05 $\pm$ 0.85 $\frac{g}{dL}$; Women 13.52 $\pm$ 0.73 $\frac{g}{dL}$ & 0.1 \\[2pt]
Lymphocytes $\times$ \textit{sex} & Men: 32.65 $\pm$ 6.84 (\%); Women 34.51 $\pm$ 6.85 (\%) & 0.1 \\[2pt]
MCH $\times$ \textit{sex} & Men: 29.30 $\pm$ 1.63 $\frac{pg}{cell}$; Women 29.40 $\pm$ 1.63 $\frac{pg}{cell}$ & 0.1 \\[2pt]
\textit{physical activity} $\times$ \textit{sex} & Active Men: 66.6\%; Active Women: 67.0\% & 43.9 \\[2pt]
Platelets $\times$ \textit{sex} & Men: 233.65 $\pm$ 46.97 $\frac{10^3}{\mu L}$; Women 252.40 $\pm$ 50.13 $\frac{10^3}{\mu L}$ & 0.1 \\[2pt]
Blood Volume Distribution $\times$ \textit{sex} & Men: 13.03 $\pm$ 0.88 (\%); Women 13.29 $\pm$ 0.93 (\%) & 0.1 \\[2pt]
WBC $\times$ \textit{sex} & Men: 5.88 $\pm$ 1.29 $\frac{10^3}{\mu L}$; Women 5.93 $\pm$ 1.38 $\frac{10^3}{\mu L}$ & 0.1 \\
\bottomrule
\end{tabular}
\caption{Covariate and interaction terms (34 total) included in Equation~\eqref{eq:lin_pred}, grouped into demographic, clinician-measured, blood-test, and interaction variables, with summary statistics computed on the training set. For numerical covariates, we report the empirical mean $\pm$ standard deviation SD; for categorical covariates, we report the empirical frequency. The last column reports the percentage of missing values (NA) for each feature. Missing values are excluded from the computation of all empirical statistics.} 
\label{tab:covariates-complete}
\end{table}
We applied the logarithmic transformation to the five target variables and to the covariates \textit{alanine ALT} and \textit{ferritin} to mitigate their right skewness (see Supplementary Figure~\ref{fig:skewed_distributions}).
Moreover, before fitting the model, we standardised the target variables and all numerical covariates.
Table~\ref{tab:target-complete} reports a summary of the five target variables in the dataset before log-transformation and standardisation.
The final set of covariates contained some missing values, although the proportion of missing observations was below 50\% for every covariate (percentages of NAs in Table~\ref{tab:covariates-complete}).
We therefore imputed the missing values using multiple imputation by chained equations, implemented through the \texttt{MICE} package in \texttt{R} \citep{van2011mice}.

\begin{table}[htbp]
\centering
\scriptsize
\begin{tabular}{lr}
\toprule
\textbf{Targets} & \textbf{Empirical Mean $\pm$ SD} \\
\midrule 
\textbf{Measured by clinician} & \\
\textit{waist circumference} & 93.81 $\pm$ 10.24 \textit{cm} \\
\textit{PMAX} & 121.83$\pm$ 11.14 \textit{mmHg} \\
\midrule[-2pt]
\textbf{Blood tests} & \\
\textit{glucose}& 91.89 $\pm$ 10.60 $\frac{mg}{dL}$ \\[2pt]
\textit{HDL cholesterol} & 56.52 $\pm$ 14.22 $\frac{mg}{dL}$ \\[2pt]
\textit{triglycerides} & 105.44 $\pm$ 55.43 $\frac{mg}{dL}$ \\
\bottomrule
\end{tabular}
\caption{Empirical means and standard deviations of the five target variables in the definition of MetS. Missing values in the \textit{waist circumference} are excluded from the computation of these statistics.}
\label{tab:target-complete}
\end{table}

See Supplementary Appendix~\ref{app:missing_cov} for more details.
The empirical prevalence of MetS in the training set was 9.2\% when considering only complete observations (i.e., excluding those with missing \textit{waist circumference}).
Across the entire training set, the overall prevalence dropped to 7.2\%.
This lower prevalence reflects a conservative estimate: for observations lacking \textit{waist circumference}, a positive MetS diagnosis is assigned only when at least three of the remaining four clinical criteria exceed their respective thresholds, thereby classifying borderline cases as negative.

\section{Bayesian multivariate modelling of longitudinal MetS biomarkers}
\label{sec:mult_GLMM}

As mentioned in the Introduction, joint modelling of the five components of MetS is crucial since they are biologically interconnected risk factors.
Modelling each component separately would overlook their inherent dependence, potentially leading to inefficient or biased inference, as noted in the context of cardiovascular diseases in \cite{RegitzZagrosek2016GenderCVD} and \cite{deBakker2025circulating}.
In particular, because we model the logarithms of the five MetS biomarkers across repeated donations with donor random effects, our model can be interpreted as a GLMM.
Let $\mathbf{Y}_{ij} = (Y_{ij}^{(1)}, \dots, Y_{ij}^{(K)})^\prime$ denote the column vector of the $K=5$ continuous outcome variables for individual $i = 1, \dots, I$ at visit $j = 1, \dots, n_i$, corresponding to the five log-transformed and standardised MetS components.
Note that the $j$th visit for the $i$th donor does not typically occur at the same time as the $j$th visit for the $i'$th donor, since donation timing depends on voluntary donor participation as well as the mandatory deferral period during which the donor is not eligible to donate again.
At each visit $j$, an associated column covariate vector $\mathbf{x}_{ij} = (x_{ij1}, \ldots, x_{ijP})^\prime \in \mathbb{R}^P$ is observed for donor $i$.
The $P$ covariates include demographic characteristics, anthropometric indices, lifestyle variables, haematological and biochemical markers, together with selected interaction terms, as reported in Table~\ref{tab:covariates-complete}.
Age at each visit is included as a time-varying covariate, thereby allowing the model to account for calendar time across donations.
We specify the following multivariate Bayesian mixed-effects model:
\begin{equation}
\label{eq:ver}
\mathbf{Y}_{ij} \mid \bm{\mu}_{ij},\, \bm{\Sigma_Y}
\ind \mathcal{N}_K(\bm{\mu}_{ij},\, \bm{\Sigma_Y}),
\quad i=1,\dots,I,\;\; j=1,\dots,n_i, 
\end{equation}
\vspace{-1cm}
\begin{equation}
\label{eq:lin_pred}
\bm{\mu}_{ij} = \bm{\beta}^\prime \mathbf{x}_{ij}+ \bm{b}_i
\end{equation}
where $\bm{\beta} := [\bm{\beta}^{(1)}, \ldots, \bm{\beta}^{(K)}] \in \mathbb{R}^{P \times K}$ is the matrix of regression coefficients, and $\bm{b}_i \in \mathbb{R}^K$ is a donor-specific random intercept vector.
Each column $\bm{\beta}^{(k)} = (\beta_{1k}, \ldots, \beta_{Pk})^\prime$ of $\bm{\beta}$ represents the regression parameters associated with $\mathbf{x}_{ij}$ on the $k$-th target outcome $Y_{ij}^{(k)}$.
Thus, each of the five target variables depends on all $P$ covariates, but with outcome-specific regression coefficients.

The covariance matrix $\bm{\Sigma_Y}$ describes the residual dependence among the five metabolic components at a given visit.
Since the target variables are (log-transformed and) standardised, $\bm{\Sigma_Y}$ represents the residual correlation matrix among the five MetS components.
We assign the following marginal prior:
\begin{equation}
\label{eq:prior_residual_variance}
\bm{\Sigma_Y} \sim IW(\nu_Y, \bm{\Psi_Y}),
\end{equation}
where $IW(\nu, \bm\Psi)$ denotes the Inverse-Wishart distribution with $\nu$ degrees of freedom and scale matrix $\bm\Psi$, such that

$\displaystyle \E(\bm{\Sigma_Y})= \bm{\Psi_{Y}} / (\nu - K - 1)$.

Regarding the regression matrix $\bm{\beta}$, to perform structured variable selection while encouraging joint sparsity across outcomes, we assign a horseshoe prior to $\bm{\beta}$.
Covariates are partitioned into three groups $C \in \{\text{Numerical}, \text{Binary}, \text{Interactions}\}$.
Thus, for each covariate index $p = 1, \dots, P$ belonging to group $C$ and each outcome $k = 1, \dots, K$, we define
\begin{equation}
\beta_{pk} = \tau_C \, \lambda_{p} \, z_{pk}
\label{eq:beta_pk}
\end{equation}
where $\{z_{pk}\}$, $\{\lambda_{p}\}$, $\{\tau_C\}$ are independent with
\begin{equation}
z_{pk} \iid \mathcal{N}(0,1), \quad
\lambda_{p} \iid \text{Cauchy}^+(0,1), \quad
\tau_C \iid\text{Cauchy}^+(0,1).
\label{eq:z_pk}
\end{equation}

Here, $\lambda_p$ is a covariate-specific local scale parameter shared across all $K$ outcomes, while $\tau_C$ is a group-specific global shrinkage parameter shared across all covariates of type $C$ and across outcomes.
Hence, the regression coefficients are conditionally independent given their respective global and local scale parameters.
This shared global--local structure induces joint shrinkage of the coefficients associated with a given covariate across the five targets, favouring predictors that are informative for multiple MetS components.
A coefficient $\beta_{pk}$ can escape substantial shrinkage only if the corresponding $\lambda_p$ is sufficiently large, thereby promoting parsimonious yet jointly predictive multivariate models.

We assume the donor-specific random intercepts $\mathbf{b}_i$ to be exchangeable and model them with the following multivariate hierarchical prior:
\begin{align}
\label{eq:prior_randomeff_param2}
\mathbf{b}_i \mid \bm{\mu}_b, \bm{\Sigma}_b &\iid \mathcal{N}_K(\bm{\mu}_b,\bm{\Sigma}_b), \;\; i = 1, \ldots,I, \;\;
\bm{\mu}_b \sim \mathcal{N}_K(\mathbf{0},\mathbf{I}_K), \; 
\bm{\Sigma}_b \sim IW(\nu_b, \bm{\Psi}_b).
\end{align}

The matrix $\bm{\Sigma}_b$ captures the covariance among donor-specific random intercepts across the five outcomes.
Formulation \eqref{eq:ver}-\eqref{eq:prior_randomeff_param2} defines a multivariate regression model with outcome-specific coefficient vectors, coupled through shared shrinkage and covariance structures.

\subsection{Posterior predictive distributions}
\label{sec:postpred}

We evaluate the predictive goodness-of-fit of our model and alternative models by computing predictions at the last visit for all donors in the training set (i.e., for all data in the validation set) and predictions for donors in the test set.
All predictions are based on the posterior predictive distribution of the five target variables.
Specifically, for each donor $i$ at the last available visit $n_i+1$, for $i=1,\ldots, I$ in the validation set, let $S_{i \ n_i+1}$ denote the indicator that the donor satisfies the MetS diagnostic criteria at visit $n_i+1$.
Given the available covariate vector ${\bm x}_{i \ n_i+1}$, our goal is to estimate the posterior predictive probability $P(S_{i\ n_i+1}=1 \mid \text{data}, {\bm x}_{i \ n_i+1})$.
This quantity is approximated via MCMC.
At each posterior draw of the model parameters, we simulate the future outcomes ${\bm Y}_{i \ n_i+1}$ according to the likelihood \eqref{eq:ver}-\eqref{eq:lin_pred} under the current values of the parameters in the MCMC, and set $S_{i \ n_i+1}=1$ if and only if at least three out of the five components meet the thresholds in Table~\ref{tab:MetS_criteria}.
The average of the resulting MCMC samples estimates the posterior predictive probability that donor $i$ will have MetS at visit $n_i+1$.
For the 689   donors in the validation set, the observed outcome $s_{i \ n_i+1}$ is also available, and it will be used to assess the goodness-of-fit in Section~\ref{sec:comparison} and also the traffic-light warning system in Section~\ref{sec:colloquium}.
Similarly, we compute the posterior predictive distribution for each donor in the test set.

For comparison across different models, we compute the Leave-One-Out Information Criterion (LOOIC) \citep{vehtari2017practical} and the Watanabe-Akaike Information Criterion (WAIC) \citep{watanabe2010asymptotic}, both of which estimate expected predictive performance while accounting for model complexity.
On the test dataset, we assess the discriminative ability of the Bayesian models using the Area Under the ROC Curve (AUC) and compare their performance with alternative machine learning (ML)  classifiers in terms of sensitivity, specificity, and overall accuracy.
Predictive performance on the test set is evaluated using metrics tailored to each modelling approach.
Because the Bayesian multivariate and autoregressive models predict continuous values for the five MetS biomarkers, their predictive accuracy is assessed using Mean Absolute Error (MAE) and Root Mean Square Error (RMSE), with MetS classification subsequently derived from the posterior predictive distributions.
In contrast, the benchmark ML models described in Section~\ref{sec:comparison} directly predict binary MetS status and therefore do not provide predictions on the original biomarker scales.
Their discriminative performance on the test set is consequently evaluated using the AUC.

\subsection{Hierarchical model selection strategy}
\label{sec:alternative_dynamic}

The final model \eqref{eq:ver}-\eqref{eq:prior_randomeff_param2} adopted in this paper is the result of a comprehensive model-building strategy to identify an appropriate representation of both the multivariate dependence among the five MetS components and their longitudinal evolution, but also the longitudinal evolution of covariates.
To identify the most appropriate modelling framework, we have followed a progressive model-building strategy aimed at answering four questions: whether multivariate modelling is preferable to independent univariate analyses; whether the current metabolic profile retains residual dependence on measurements from the previous donation; whether the effect of time-varying covariates should be modelled dynamically over the inter-donation interval; and, ultimately, which model provides the best compromise between predictive performance, parsimony, computational efficiency, and clinical interpretability.
Overall, our multivariate mixed-effects model \eqref{eq:ver}-\eqref{eq:prior_randomeff_param2} provided the best overall compromise across these criteria, indicating that the multivariate dependence structure and donor-specific random effects at the same visit adequately capture the longitudinal evolution of the MetS biomarkers without requiring explicit autoregressive components.
See Supplementary Appendix~\ref{sec:appendixB} for details.
In particular, Supplementary Table~\ref{tab:full_model_comparison} reports information criteria and predictive performance metrics across all the Bayesian multivariate models fitted to the data.
The increased complexity of the autoregressive specifications also comes at a substantial computational cost, with fitting times approximately twice those required by the Bayesian multivariate model in \eqref{eq:ver}-\eqref{eq:prior_randomeff_param2}.
In all cases, posterior estimates of the t-CAR decay parameters (reported in Supplementary Appendix~\ref{sec:appendixB}) suggest that temporal autocorrelation drops to virtually zero over the typical time intervals between consecutive blood donations.
Furthermore, model comparison on the    test set in Supplementary Appendix~\ref{app:model_selection} shows that introducing explicit autoregressive components provides no meaningful predictive improvement over the likelihood specification \eqref{eq:ver}-\eqref{eq:lin_pred}.
The comparison also supports the joint multivariate framework over independent univariate models.
By borrowing strength across correlated metabolic outcomes, the multivariate structure reduces uncertainty when imputing missing outcome values.
Taken together, these results, along with the fact that age interactions already partially capture longitudinal metabolic dynamics, support excluding explicit autoregressive components from the final model.

\section{Posterior inference}
\label{sec:results}

We compute the posterior distribution of the Bayesian model \eqref{eq:ver}-\eqref{eq:prior_randomeff_param2} using an MCMC algorithm based on the training set.
In particular, we employ Hamiltonian Monte Carlo through the software platform Stan \citep{carpenter2017stan}.
We fix moderately informative hyperparameters, 
$\nu_Y = \nu_b = 10$, $\bm{\Psi}_Y = 2 I$, and $\bm{\Psi}_b = 12 I$, 
so that $\E(\bm{\Sigma}_Y) = 0.5 I$ and $\E(\bm{\Sigma}_b) = 3 I$.
Under this specification, the prior variances are 
$\Var(\Sigma_{Y,ii}) = 0.25$ and $\Var(\Sigma_{b,ii}) = 9$ for the diagonal elements, and
$\Var(\Sigma_{Y,ij}) \approx 0.167$ and $\Var(\Sigma_{b,ij}) = 6$ for the off-diagonal elements
($i \neq j$).
Since the target variables are standardised before modelling, this prior maintains a relatively small expected residual variance for within-donor fluctuations ($\bm{\Sigma}_Y$), while acting as a permissive, weakly informative prior for the random effects ($\bm{\Sigma}_b$), allowing the model to capture substantial baseline heterogeneity across donors freely.
We run four parallel MCMC chains with 1,000 burn-in and 1,000 sampling iterations each, for a total of 4,000 draws.
Convergence of the MCMC algorithm has been assessed: no divergent transitions were reported during sampling 
and the E-BFMI diagnostic metric indicates no pathological behaviour in the energy exploration.
We have also performed a sensitivity analysis on the prior hyperparameters of the random-effects covariance matrix, namely $\bm{\Psi}_b$ and $\nu_b$, and obtained consistent posterior results across different prior specifications.
 
\begin{table}[htbp]
\centering
\tiny
\begin{tabular}{llllll}
\toprule
Covariate & glucose & HDL & PMAX & triglycerides & waist circumference \\
\midrule
ALT & \textcolor{black}{0.03 , (-0.00, 0.05)} & \textcolor{black}{0.01 , (-0.01, 0.02)} & \textcolor{black}{-0.01 , (-0.03, 0.02)} & \textcolor{black}{0.02 , (-0.00, 0.04)} & \textcolor{black}{0.01 , (-0.00, 0.03)} \\
Age & \textcolor{darkgreen}{0.13 , (0.08, 0.17)} & \textcolor{darkgreen}{0.05 , (0.00, 0.10)} & \textcolor{darkgreen}{0.31 , (0.23, 0.35)} & \textcolor{darkgreen}{0.15 , (0.11, 0.20)} & \textcolor{darkgreen}{0.20 , (0.14, 0.23)} \\
BMI & \textcolor{darkgreen}{0.10 , (0.04, 0.14)} & \textcolor{darkred}{-0.28 , (-0.32, -0.24)} & \textcolor{darkgreen}{0.22 , (0.17, 0.25)} & \textcolor{darkgreen}{0.21 , (0.16, 0.25)} & \textcolor{darkgreen}{0.59 , (0.56, 0.63)} \\
Basophils\_pct & \textcolor{black}{0.00 , (-0.02, 0.02)} & \textcolor{black}{0.01 , (-0.00, 0.03)} & \textcolor{black}{0.00 , (-0.01, 0.02)} & \textcolor{black}{-0.01 , (-0.03, 0.01)} & \textcolor{black}{-0.00 , (-0.02, 0.01)} \\
Creatinine & \textcolor{black}{-0.00 , (-0.02, 0.02)} & \textcolor{black}{-0.01 , (-0.02, 0.01)} & \textcolor{black}{-0.01 , (-0.03, 0.01)} & \textcolor{black}{0.01 , (-0.01, 0.04)} & \textcolor{black}{-0.01 , (-0.03, 0.00)} \\
\addlinespace
Eosinophils\_pct & \textcolor{black}{0.00 , (-0.02, 0.03)} & \textcolor{darkred}{-0.03 , (-0.05, -0.01)} & \textcolor{black}{-0.01 , (-0.03, 0.01)} & \textcolor{darkgreen}{0.05 , (0.02, 0.07)} & \textcolor{black}{0.00 , (-0.01, 0.02)} \\
Ferritin & \textcolor{black}{0.00 , (-0.02, 0.03)} & \textcolor{black}{-0.02 , (-0.04, 0.00)} & \textcolor{black}{0.01 , (-0.02, 0.03)} & \textcolor{black}{-0.01 , (-0.03, 0.01)} & \textcolor{black}{0.01 , (-0.01, 0.02)} \\
Heart\_rate & \textcolor{black}{0.01 , (-0.02, 0.04)} & \textcolor{black}{-0.01 , (-0.03, 0.01)} & \textcolor{black}{0.06 , (-0.00, 0.10)} & \textcolor{black}{0.03 , (-0.01, 0.06)} & \textcolor{black}{0.00 , (-0.03, 0.02)} \\
Height & \textcolor{black}{0.00 , (-0.04, 0.04)} & \textcolor{black}{-0.03 , (-0.08, 0.02)} & \textcolor{black}{0.01 , (-0.04, 0.05)} & \textcolor{black}{-0.01 , (-0.06, 0.04)} & \textcolor{darkgreen}{0.22 , (0.18, 0.25)} \\
Hemoglobin & \textcolor{black}{-0.00 , (-0.04, 0.03)} & \textcolor{darkgreen}{0.11 , (0.08, 0.14)} & \textcolor{darkgreen}{0.07 , (0.03, 0.10)} & \textcolor{darkgreen}{0.13 , (0.08, 0.16)} & \textcolor{black}{-0.00 , (-0.03, 0.03)} \\
\addlinespace
Lymphocytes \% & \textcolor{black}{-0.01 , (-0.04, 0.02)} & \textcolor{black}{-0.01 , (-0.03, 0.01)} & \textcolor{black}{-0.03 , (-0.06, 0.01)} & \textcolor{black}{0.03 , (-0.00, 0.06)} & \textcolor{black}{-0.00 , (-0.02, 0.02)} \\
MCH & \textcolor{black}{0.04 , (-0.01, 0.07)} & \textcolor{darkgreen}{0.12 , (0.07, 0.15)} & \textcolor{darkred}{-0.08 , (-0.12, -0.04)} & \textcolor{darkred}{-0.06 , (-0.09, -0.01)} & \textcolor{darkred}{-0.04 , (-0.06, -0.01)} \\
MCHC & \textcolor{darkred}{-0.08 , (-0.11, -0.05)} & \textcolor{darkred}{-0.10 , (-0.12, -0.08)} & \textcolor{black}{0.02 , (-0.01, 0.05)} & \textcolor{black}{0.02 , (-0.01, 0.05)} & \textcolor{black}{0.01 , (-0.01, 0.02)} \\
Monocytes \% & \textcolor{darkred}{-0.09 , (-0.12, -0.06)} & \textcolor{black}{0.00 , (-0.01, 0.02)} & \textcolor{darkred}{-0.04 , (-0.07, -0.02)} & \textcolor{darkred}{-0.02 , (-0.05, -0.00)} & \textcolor{darkgreen}{0.01 , (0.00, 0.03)} \\
Platelets & \textcolor{black}{-0.00 , (-0.03, 0.02)} & \textcolor{black}{0.02 , (-0.00, 0.05)} & \textcolor{black}{-0.00 , (-0.03, 0.02)} & \textcolor{black}{0.01 , (-0.01, 0.03)} & \textcolor{black}{0.00 , (-0.01, 0.02)} \\
\addlinespace
Total\_proteins & \textcolor{black}{0.02 , (-0.00, 0.04)} & \textcolor{darkgreen}{0.02 , (0.01, 0.03)} & \textcolor{black}{-0.00 , (-0.02, 0.01)} & \textcolor{darkgreen}{0.02 , (0.00, 0.03)} & \textcolor{black}{0.00 , (-0.01, 0.01)} \\
Volume\_distribution & \textcolor{black}{0.00 , (-0.03, 0.03)} & \textcolor{darkgreen}{0.07 , (0.04, 0.09)} & \textcolor{black}{-0.00 , (-0.03, 0.03)} & \textcolor{black}{-0.02 , (-0.05, 0.00)} & \textcolor{black}{-0.01 , (-0.03, 0.01)} \\
WBC & \textcolor{black}{-0.01 , (-0.05, 0.03)} & \textcolor{black}{0.01 , (-0.02, 0.05)} & \textcolor{black}{-0.03 , (-0.07, 0.01)} & \textcolor{darkgreen}{0.09 , (0.02, 0.14)} & \textcolor{black}{0.01 , (-0.02, 0.03)} \\
Physical\_activity\_Active & \textcolor{black}{-0.04 , (-0.09, 0.01)} & \textcolor{black}{-0.00 , (-0.03, 0.02)} & \textcolor{black}{0.01 , (-0.04, 0.05)} & \textcolor{black}{0.01 , (-0.03, 0.05)} & \textcolor{darkred}{-0.11 , (-0.15, -0.06)} \\
\textit{sex}\_Male & \textcolor{darkgreen}{0.15 , (0.03, 0.28)} & \textcolor{darkred}{-0.87 , (-1.01, -0.73)} & \textcolor{darkgreen}{0.26 , (0.13, 0.39)} & \textcolor{darkgreen}{0.16 , (0.02, 0.30)} & \textcolor{darkgreen}{0.37 , (0.27, 0.46)} \\
\addlinespace
ALT $\times$ \textit{sex} & \textcolor{black}{0.01 , (-0.01, 0.04)} & \textcolor{black}{0.00 , (-0.01, 0.02)} & \textcolor{black}{0.00 , (-0.02, 0.02)} & \textcolor{black}{0.01 , (-0.01, 0.04)} & \textcolor{black}{-0.00 , (-0.03, 0.01)} \\
Age $\times$ \textit{sex} & \textcolor{black}{0.01 , (-0.02, 0.07)} & \textcolor{black}{0.00 , (-0.03, 0.05)} & \textcolor{black}{0.02 , (-0.01, 0.11)} & \textcolor{black}{-0.00 , (-0.05, 0.03)} & \textcolor{black}{0.01 , (-0.01, 0.07)} \\
BMI $\times$ \textit{sex} & \textcolor{black}{0.01 , (-0.02, 0.07)} & \textcolor{black}{-0.01 , (-0.05, 0.02)} & \textcolor{black}{0.01 , (-0.02, 0.05)} & \textcolor{black}{0.01 , (-0.02, 0.06)} & \textcolor{black}{-0.01 , (-0.06, 0.01)} \\
Creatinine $\times$ \textit{sex} & \textcolor{black}{-0.00 , (-0.02, 0.01)} & \textcolor{black}{-0.00 , (-0.02, 0.01)} & \textcolor{black}{-0.00 , (-0.02, 0.01)} & \textcolor{black}{-0.00 , (-0.02, 0.01)} & \textcolor{black}{0.00 , (-0.01, 0.01)} \\
Ferritin $\times$ \textit{sex} & \textcolor{black}{-0.00 , (-0.02, 0.02)} & \textcolor{black}{-0.01 , (-0.03, 0.01)} & \textcolor{black}{0.00 , (-0.01, 0.02)} & \textcolor{black}{-0.00 , (-0.03, 0.01)} & \textcolor{black}{0.00 , (-0.01, 0.02)} \\
\addlinespace
Hb $\times$ \textit{sex} & \textcolor{black}{0.00 , (-0.02, 0.03)} & \textcolor{black}{0.00 , (-0.01, 0.03)} & \textcolor{black}{0.00 , (-0.02, 0.03)} & \textcolor{black}{0.01 , (-0.01, 0.06)} & \textcolor{black}{-0.00 , (-0.03, 0.02)} \\
Heart rate$\times$ \textit{sex} & \textcolor{black}{0.01 , (-0.01, 0.04)} & \textcolor{black}{-0.01 , (-0.03, 0.01)} & \textcolor{black}{0.03 , (-0.01, 0.10)} & \textcolor{black}{0.01 , (-0.01, 0.06)} & \textcolor{black}{0.01 , (-0.01, 0.04)} \\
Height $\times$ \textit{sex} & \textcolor{black}{-0.00 , (-0.03, 0.03)} & \textcolor{black}{-0.00 , (-0.03, 0.02)} & \textcolor{black}{-0.00 , (-0.03, 0.02)} & \textcolor{black}{-0.00 , (-0.03, 0.02)} & \textcolor{black}{0.00 , (-0.02, 0.03)} \\
Lymphocytes \% $\times$ \textit{sex} & \textcolor{black}{-0.01 , (-0.05, 0.01)} & \textcolor{black}{-0.00 , (-0.03, 0.01)} & \textcolor{black}{-0.01 , (-0.05, 0.01)} & \textcolor{black}{0.00 , (-0.02, 0.04)} & \textcolor{black}{0.00 , (-0.01, 0.03)} \\
Mch $\times$ \textit{sex} & \textcolor{black}{0.01 , (-0.01, 0.05)} & \textcolor{black}{0.01 , (-0.01, 0.06)} & \textcolor{black}{-0.00 , (-0.04, 0.02)} & \textcolor{black}{-0.01 , (-0.05, 0.01)} & \textcolor{black}{-0.00 , (-0.03, 0.01)} \\
\addlinespace
PhAct $\times$ \textit{sex} & \textcolor{black}{0.00 , (-0.02, 0.03)} & \textcolor{black}{0.00 , (-0.02, 0.02)} & \textcolor{black}{0.00 , (-0.01, 0.04)} & \textcolor{black}{0.00 , (-0.02, 0.03)} & \textcolor{black}{-0.00 , (-0.04, 0.02)} \\
Piastrine $\times$ \textit{sex} & \textcolor{black}{-0.00 , (-0.02, 0.02)} & \textcolor{black}{0.01 , (-0.01, 0.04)} & \textcolor{black}{0.00 , (-0.01, 0.02)} & \textcolor{black}{0.00 , (-0.02, 0.02)} & \textcolor{black}{0.00 , (-0.01, 0.02)} \\
Volume$\times$ \textit{sex} & \textcolor{black}{-0.00 , (-0.03, 0.01)} & \textcolor{black}{0.00 , (-0.01, 0.03)} & \textcolor{black}{0.00 , (-0.01, 0.03)} & \textcolor{black}{-0.00 , (-0.02, 0.02)} & \textcolor{black}{0.00 , (-0.02, 0.02)} \\
WBC $\times$ \textit{sex} & \textcolor{black}{-0.02 , (-0.07, 0.01)} & \textcolor{black}{-0.02 , (-0.06, 0.01)} & \textcolor{black}{0.00 , (-0.04, 0.04)} & \textcolor{black}{0.04 , (-0.00, 0.12)} & \textcolor{black}{0.01 , (-0.01, 0.05)} \\
\bottomrule
\end{tabular}
\caption{Under the horseshoe prior specification, posterior means and 95\% credible intervals of the fixed effects for each covariate across the five target variables. Green and red colors denote positive and negative coefficients whose 95\% credible intervals exclude zero, respectively.}
\label{tab:coeff_post_estimate}
\end{table}
Our Bayesian hierarchical multivariate mixed-effects model provides robust and interpretable inferences on individual MetS risk within the AVIS Milan donor population.
Table~\ref{tab:coeff_post_estimate} shows that only a small subset of predictors remains clearly associated with the MetS component variables, while most coefficients are shrunk toward zero by the horseshoe prior.
The posterior means of the fixed effects are consistent with established clinical knowledge: body mass index (\textit{BMI}), \textit{age}, and \textit{sex} emerge as central determinants across the five MetS biomarker outcomes.
Not surprisingly, \textit{BMI} shows strong positive associations with \textit{PMAX}, \textit{triglycerides}, \textit{waist circumference}, and \textit{glucose}, and a strong negative association with \textit{HDL cholesterol}.
Male \textit{sex} is negatively correlated with \textit{HDL cholesterol}, confirming that men physiologically tend to have lower \textit{HDL cholesterol} levels, and is positively correlated with all the other target variables.
These findings are consistent with known \textit{sex} differences in lipid metabolism and blood pressure regulation \citep{RegitzZagrosek2016GenderCVD}.
\textit{Age} is one of the main drivers of all five target variables.

The non-zero coefficients for \textit{HDL cholesterol} associated with the two red blood cell indices, \textit{MCH} and \textit{MCHC}, which relate to haemoglobin content and concentration, suggest that these indices may capture additional information about lipid metabolism beyond traditional MetS markers.
Higher or lower \textit{MCH}/\textit{MCHC} values often reflect subtle alterations in iron status, chronic inflammation, or nutritional deficiencies, all of which have been associated with atherogenic dyslipidemia and variations in \textit{HDL} levels \citep{weiss2019anemia,krawiec2020biomarkers,clucas2022interpretation}.
Their significance in our model supports the notion that low-grade inflammatory or haematologic changes may coexist with the dyslipidemic profile typical of MetS.
Note that none of the interaction terms between \textit{sex} and other covariates is statistically relevant.

\begin{figure*}[htbp]
\centering

  \includegraphics[width=\textwidth]{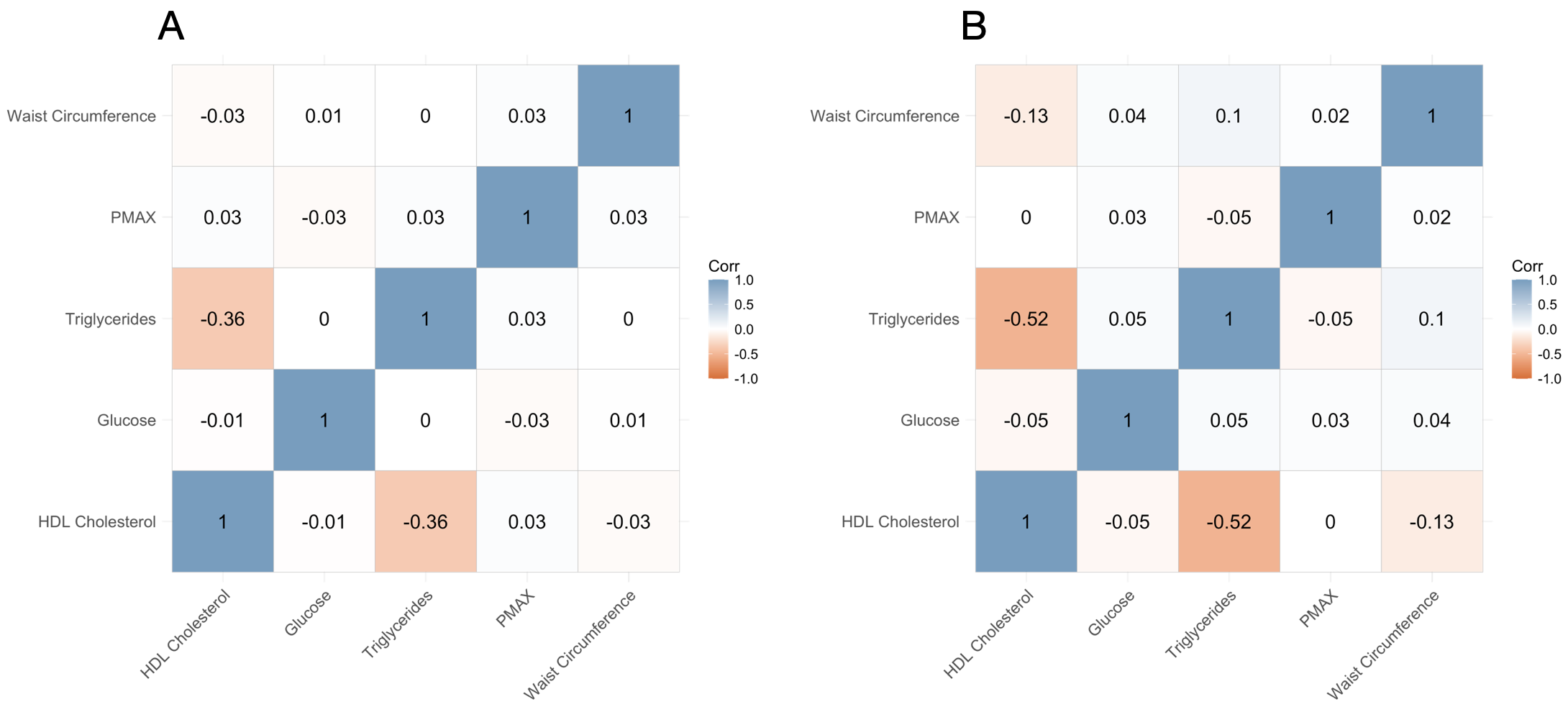}
  \vspace{4pt}

  \begin{minipage}[t]{0.48\textwidth}
    \centering
    \small (A) Residuals ($\bm{\Sigma_Y}$)
    \label{fig:residual_correlations}
  \end{minipage}\hfill
  \begin{minipage}[t]{0.48\textwidth}
    \centering
    \small (B) Random effects ($\bm{\Sigma_b}$)
    \label{fig:random_correlations}
  \end{minipage}

\vspace{6pt}
\caption{Posterior mean of the covariance matrices $\Sigma_Y$ and $\Sigma_b$.}
\label{fig:correlation_matrices}
\end{figure*}

Our hierarchical Bayesian model explicitly accounts for correlations among the five target variables by modelling a full covariance structure for both the random intercepts ($\bm{\Sigma_b}$) and the residuals ($\bm{\Sigma_Y}$).
Figure~\ref{fig:correlation_matrices} shows the posterior means of the covariance matrices $\bm{\Sigma_Y}$ and $\bm{\Sigma_b}$, revealing predominantly weak associations.
The strongest residual covariance is observed between \textit{HDL cholesterol} and \textit{triglycerides} ($-0.36$) (Figure~\ref{fig:correlation_matrices}, panel~a).
Panel~(b) shows stronger covariances at the random intercept level, including \textit{triglycerides}-\textit{HDL-cholesterol} ($-0.52$), \textit{waist circumference}-\textit{HDL-cholesterol} ($-0.13$), and \textit{waist circumference}-\textit{triglycerides} ($0.1$).
These results offer important insights into pathophysiology and modelling.
The moderate negative covariance between \textit{HDL cholesterol} and \textit{triglycerides}, present at both the residual and subject-specific levels, reflects the well-documented biological inverse relationship between these two lipids, a hallmark of atherogenic dyslipidemia and insulin resistance \citep{rashid2003mechanisms}.
Crucially, the larger covariances among the random intercepts ($\bm{\Sigma_b}$) than among the residuals ($\bm{\Sigma_Y}$) indicate that the underlying metabolic dependence is primarily driven by unobserved, time-invariant individual characteristics (e.g., genetic predispositions or long-term lifestyle habits not fully captured by covariates) rather than short-term fluctuations.
These non-negligible associations support modelling the dependence structure among target variables rather than treating them as independent.

\section{Comparison with alternative classification methods and simulation study}
\label{sec:comparison}

In this section, we present the comparison between the Bayesian classification rule introduced in Section~\ref{sec:postpred} and several widely used ML classification approaches: Logistic Regression (LR), Linear Discriminant Analysis (LDA), K-Nearest Neighbors (KNN), and Light Gradient Boosting Machine (LGBM) \citep{ke2017lightgbm}, assuming independence over time.
LR, LDA, and KNN are standard statistical techniques widely used in biomedical research, providing transparent parametric (LR, LDA) or distance-based (KNN) decision rules that are relatively easy to interpret.
In contrast, LGBM is a modern ML method based on gradient-boosted decision trees, designed to achieve high predictive accuracy on tabular data by accommodating nonlinear effects and complex interactions, albeit at the cost of reduced interpretability. We compared these methods on both the   validation set and test set.

We fit these competing approaches using complete observations from the training set, resulting in a subset of 5,118 fully observed visits from 2,177 donors; 
donations with missing \textit{waist circumference} were excluded, as these methods cannot automatically handle missing values in the response.
All approaches used the same covariates as our Bayesian model in \eqref{eq:lin_pred}, as reported in Table~\ref{tab:covariates-complete}.
The competing approaches  {have been fitted} in \texttt{R}.
LR has been fitted with \texttt{glmnet} \citep{friedman2010regularization}, setting $\alpha=1$ to impose a LASSO penalty on the coefficients and selecting the optimal regularisation parameter $\lambda$ via 5-fold cross-validation.
LDA  {has been} fitted with the \texttt{MASS} package and requires no hyperparameter tuning.
KNN  {has been} implemented with the \texttt{class} package, with the optimal number of neighbors $k$ selected via leave-one-out cross-validation.
Finally, LGBM  {has been} fitted with \texttt{lightgbm} v4.6.0 \citep{lightgbmR}, with hyperparameters tuned via Bayesian optimisation using \texttt{rBayesianOptimization} \citep{yan_rbayesianoptimization}.

For each classification method, we  {have} selected a binary classification threshold $t$ by maximising a weighted version of Youden's $J_r$ statistic on the training set, defined as 
\begin{equation}
 J_r=\text{sensitivity}+r\cdot\text{specificity}-1, 
 \label{eq:Youdens}
\end{equation}
where $r$ controls the relative weight assigned to specificity versus sensitivity.
In this context, sensitivity (or true positive rate) is the proportion of MetS-positive donors correctly identified as such by the model, while specificity (true negative rate) is the proportion of MetS-negative (healthy) donors correctly identified as such by the model.
We also evaluated accuracy, which is defined as the total proportion of correctly classified donations (i.e., the sum of true positives and true negatives divided by the total number of evaluated donations).

A donor  {is} classified as MetS-negative if the estimated probability of MetS is below $t$, and as MetS-positive otherwise.
We set $r=0.75$ in \eqref{eq:Youdens} to prioritise sensitivity, consistent with the clinical premise that false-negative classifications (failure to identify an at-risk donor) entail more serious consequences than false-positive classifications (unnecessarily assigning a healthy donor to enhanced monitoring).

Note that, when using the standard unweighted Youden's index ($r=1$), all models yielded lower values of sensitivity, in favour of higher specificity.
The thresholds obtained by maximising $J_r$ are: $t=0.072$ for our Bayesian model; $t=0.047$ for LR, $t=0.067$ for LDA, $t=0.113$ for KNN, and $t=0.635$ for LGBM.
We  {have} additionally evaluated discrimination independently of the selected threshold by computing the AUC on the    test set for all methods.
\begin{table}[htbp]
\centering
\begin{tabular}{llcccccc}
\toprule
 && \multicolumn{5}{c}{\textbf{Validation set}} & \textbf{Test set} \\
\cmidrule(lr){3-7} \cmidrule(lr){8-8}
 && \multicolumn{2}{c}{\textbf{Confusion Matrix}} & \multicolumn{3}{c}{\textbf{Performance Metrics}} & \\
\cmidrule(lr){3-4} \cmidrule(lr){5-7}
\textbf{Model} && \textbf{True No} & \textbf{True Yes} & \textbf{Sens} & \textbf{Spec} & \textbf{Acc} & \textbf{AUC} \\
\midrule
\textbf{Bayesian} & Pred No & 527 & 3 & 94.6\% & 83.1\% & 84.0\% & 0.887 \\
\textbf{binary} & Pred Yes & 107 & 52 &&&& \\
\midrule
\textbf{LR} & Pred No & 440 & 7 & 87.3\% & 69.4\% & 70.8\% & 0.920 \\
 & Pred Yes & 194 & 48 &&&& \\
\midrule
\textbf{LDA} & Pred No & 299 & 4 & 93.5\% & 47.7\% & 51.8\% & 0.858 \\
 & Pred Yes & 328 & 58 &&&& \\
\midrule
\textbf{KNN} & Pred No & 449 & 19 & 69.4\% & 71.6\% & 71.4\% & 0.811 \\
 & Pred Yes & 178 & 43 &&&& \\
\midrule
\textbf{LGBM} & Pred No & 481 & 9 & 83.6\% & 75.8\% & 76.5\% & 0.883 \\
 & Pred Yes & 153 & 46 &&&& \\
\bottomrule
\end{tabular}
\caption{Confusion matrix and predictive performance metrics of the donors under different models/classifiers. Metrics are categorised by their evaluation data subset: confusion matrix along with sensitivity, specificity, and accuracy are evaluated on the validation set, while the AUC is evaluated on the test set.}
\label{tab:cm_all_models}
\end{table}
Table~\ref{tab:cm_all_models} reports the confusion matrices together with sensitivity, specificity, and accuracy on the   validation set, and the AUC on the    test set.
The Bayesian binary classification rule described in Section~\ref{sec:postpred} shows strong predictive performance on the   validation set, correctly identifying 94.6\% of MetS-positive
donors while maintaining high specificity (83.1\%) and accuracy (84.0\%).
These results indicate that the underlying multivariate framework effectively discriminates between affected and healthy profiles.
On the test set, the Bayesian model maintained good discriminative performance (AUC $=0.887$), comparable to LGBM (AUC $=0.883$), while LR achieves a slightly higher AUC ($0.920$).

A key advantage of the Bayesian approach lies in its generative hierarchical formulation.
The model specifies the joint distribution of the five MetS biomarkers conditional on covariates, explicitly accounting for longitudinal dependence and cross-outcome correlations.
Classification is then derived from the posterior predictive distribution.
In contrast, LR, LDA, KNN, and LGBM are discriminative classifiers that directly target class membership without modelling the underlying multivariate biomarker structure.
These competing methods, however, have a clear computational advantage: model fitting requires only a few minutes, whereas running the MCMC algorithm for our Bayesian model takes approximately three hours on a MacBook Pro with an M4 Pro chip and 24 GB of RAM.

We further assess the robustness of our model through a simulation study that replicates the longitudinal structure of the AVIS Milan cohort while allowing for model specification and temporal dependence.
Specifically, we  {have simulated} data with autoregressive dynamics and heavy-tailed noise, neither of which was explicitly accounted for in the likelihood in \eqref{eq:ver} and \eqref{eq:lin_pred}.
Despite these departures from the model assumptions, our model still yields accurate predictions on the test set.
Full details on the simulation design and posterior predictive evaluation are provided in Supplementary Appendix~\ref{sec:appendixC}.

\section{Pre-donation medical consultation}
\label{sec:colloquium}

We briefly summarise the donation procedure here.
Upon arrival, each donor registers at the reception desk, where staff verifies that the mandatory interval since the previous donation has elapsed.
Before the medical examination, donors complete a health and lifestyle questionnaire and undergo a quick finger-prick haemoglobin test, with results available immediately.
Then, a clinician performs a pre-donation visit, which is typically very short and mostly consists of a conversation.
During this assessment, the clinician reviews the donor's questionnaire, previous laboratory results (typically from the last donation or the most recent annual health check-up), and the current haemoglobin level.
Based on this information, the physician determines whether the donor is eligible to donate at the current time.
If the donor is deemed eligible, the donation proceeds, and blood samples are collected simultaneously for the mandatory laboratory analyses.
Additional tests, such as the annual health check-up or
further clinically indicated examinations, may also be performed.
Laboratory test results are made available within a few days and are sent to the donor.

For donor $i$, the posterior predictive probability of meeting the MetS diagnostic criteria at visit $n_i+1$
is computed as described in Section~\ref{sec:postpred}.
Here, $n_i+1$ denotes any future visit; it can occur in the near future, such as the current pre-donation visit $n_i+1$ where no current blood test result is available yet, or at an arbitrarily distant time point.
For any such future visit, our framework computes the posterior predictive probability of MetS, conditional on specified values of selected covariates, while {holding the remaining covariates at} their most recently observed values.
{This probability varies over time, which is included as the covariate \textit{age} in the model}. 
Thus, the model is not limited to predictions made immediately before laboratory results become available; it also allows clinicians to evaluate hypothetical lifestyle scenarios by assigning alternative values to selected modifiable covariates, such as BMI, and comparing the resulting predicted risks.

To facilitate their use,
the individual posterior predictive probabilities of MetS are discretised into a traffic-light warning system that stratifies donors into three risk levels: low (green), intermediate (yellow) and high (red).
Building on the primary decision boundary established in Section~\ref{sec:comparison}, we set the threshold $t_2$ separating the yellow and red categories to $0.072$ (via maximisation of $J_{0.75}$).
To identify a low-risk green group while limiting false negatives, we {have evaluated} a grid of alternative $r$ values in \eqref{eq:Youdens}.
Table~\ref{tab:r_youdens} reports the resulting thresholds, sensitivities, and specificities obtained on the training set.
We select $t_1=0.021$, corresponding to $r=0.20$, as the green--yellow boundary, since lower values of $r$ lead to a noticeable decrease in specificity with only a negligible improvement in sensitivity.

\begin{table}[htbp]
 \centering
 \begin{tabular}{cccc}
 \toprule
 Parameter $r$ & Threshold & Sensitivity (\%) & Specificity (\%) \\
 \midrule
 0.10 & 0.015 & 98.46 & 67.97 \\
 0.20 & 0.021 & 98.11 & 71.32 \\
 0.30 & 0.022 & 97.94 & 72.04 \\
 0.40 & 0.031 & 96.91 & 75.24 \\
 0.50 & 0.053 & 94.68 & 80.59 \\
 0.75 & 0.072 & 92.62 & 83.62 \\
 1.00 & 0.114 & 88.68 & 88.21 \\
 \bottomrule
 \end{tabular}
\caption{Threshold selected by maximizing $J_r$ in~\eqref{eq:Youdens}, with the corresponding training-set sensitivity and specificity, for different values of $r$.} 
\label{tab:r_youdens}
\end{table}

Summing up, donors are classified as green (low risk) when the individual posterior predictive probability of meeting the MetS criteria is below $t_1=0.021$, yellow (intermediate risk) for probabilities between $t_1=0.021$ and $t_2=0.072$, and red (high risk) for probabilities at or above $t_2=0.072$.
The yellow category provides an additional precautionary warning for donors whose predicted risk is insufficient for classification in the red category but who may nevertheless warrant closer monitoring.

The classification performance {has been} evaluated on the validation set against a ground truth available for 689 donors with complete target-variable measurements.
Table~\ref{tab:trafficlight_classification_testset} reports the confusion matrix for the three traffic-light categories.

\begin{table}[htbp]
\centering
\begin{tabular}{llcc}
\toprule
&& \multicolumn{2}{c}{\textbf{Confusion Matrix}} \\
\cmidrule(lr){3-4}
&& \textbf{True No} & \textbf{True Yes} \\
\midrule
\textbf{Traffic-light algorithm} & Green & 500 & 2 \\
 & Yellow & 27 & 1 \\
 & Red & 107 & 52 \\
\bottomrule
\end{tabular}
\caption{Confusion matrix of the validation set donors stratified by the proposed traffic-light warning system.}
\label{tab:trafficlight_classification_testset}
\end{table}

Because the overall time the AVIS clinician can devote to each donor is very limited, the traffic-light system could help identify donors who may require more attention during the pre-donation consultation.
In particular, clinicians could spend more time with donors classified as red or yellow, balancing the generally short duration of donor visits with the greater attention required by donors at higher predicted risk.

We are going to provide AVIS Milan with a software application that uses covariate values available at the current visit to estimate the donor's traffic-light category.
The application will also evaluate hypothetical scenarios for a subsequent visit by varying selected modifiable covariates while holding the remaining covariates fixed.
For example, the clinician will be able to show how the predicted category changes if BMI remains at its most recently observed value, decreases or increases by a specified percentage, or reaches a specified target value.
This can inform donors of their predicted risk and of the set of modifiable covariates identified as relevant in the analysis in Section~\ref{sec:results}.

\begin{figure*}[htbp]
\centering

  \includegraphics[width=\textwidth]{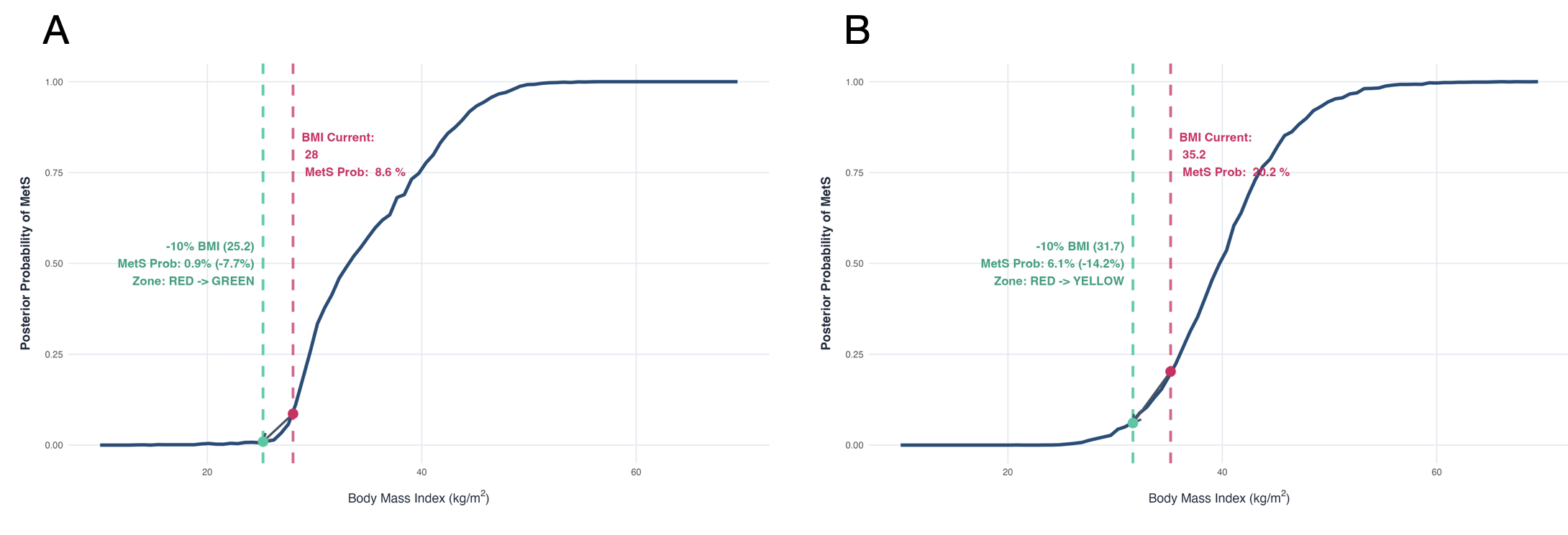}
  \vspace{4pt}

  \begin{minipage}[t]{0.48\textwidth}
    \centering
    \small (A) BMI $28\,\mathrm{kg/m^2}$
  \end{minipage}\hfill
  \begin{minipage}[t]{0.48\textwidth}
    \centering
    \small (B) BMI $35.2\,\mathrm{kg/m^2}$
  \end{minipage}

\vspace{6pt}
\caption{Posterior predictive probability of MetS as a function of BMI for two representative high-risk donors: (A) BMI \(28\,\mathrm{kg/m^2}\) and (B) BMI \(35.2\,\mathrm{kg/m^2}\). Solid lines show the predicted probabilities, while dashed lines indicate the current BMI (red) and a hypothetical 10\% BMI reduction (green), with all other covariates held fixed.}
\label{fig:bmi_intervention}
\end{figure*}

To concretely illustrate the clinical utility of our system during the pre-donation assessment, we {have simulated}  a potential personalised intervention scenario focused on weight reduction.
By incorporating subject-specific random effects and longitudinal covariates, the Bayesian model goes beyond static classification, allowing clinicians to trace dynamic, individualised risk trajectories.
As an example, Figure~\ref{fig:bmi_intervention} illustrates the simulated impact of a 10\% reduction in BMI for two donors with different obesity profiles.

The first case (Figure~\ref{fig:bmi_intervention}a) examines a 53-year-old male donor who, despite maintaining a physically active lifestyle, is currently overweight with a BMI of 28.0 kg/m$^2$ (height of 177.7 cm).
His covariates at the considered donation include haemoglobin level of 15.55 g/dL, total proteins at 7.59 g/dL, white blood cell (WBC) count of 5.39 $\times 10^3/\mu$L, alongside standard red blood cell indices (MCH: 28.11 pg, MCHC: 32.06 g/dL, red cell distribution width: 13.00\%) and normal leukocyte differentials (monocytes: 9.54\%, eosinophils: 3.06\%).
Based on this profile, the model estimates his MetS posterior probability at 6 months equal to 8.6\% if all the covariates do not change.
Because this value exceeds the $t_2=0.072$ critical threshold established in Section~\ref{sec:comparison}, our traffic-light algorithm flags the donor as red (high risk).
During the current pre-donation assessment, the clinician might visually demonstrate to the donor that a realistic, targeted goal, such as a 10\% reduction in BMI (decreasing it to 25.2 kg/m$^2$), would reduce his MetS posterior probability to 0.9\%, keeping fixed all the other covariates.
This weight loss would shift the donor's risk profile into the Green zone.

The second case (Figure~\ref{fig:bmi_intervention}b) examines a 48-year-old woman who leads a sedentary lifestyle and presents with severe obesity, with a BMI of 35.2 kg/m$^2$ (height of 150.9 cm).
Her clinical status at the time of the visit indicates a haemoglobin level of 13.23 g/dL, total proteins at 6.90 g/dL, and WBC count of 8.18 $\times 10^3/\mu$L.
Her red blood cell indices (MCH: 28.51 pg, MCHC: 31.14 g/dL, red cell distribution width: 13.20\%) and leukocyte differentials (monocytes: 5.82\%, eosinophils: 0.68\%) complete her profile.
Driven by these covariates, the model estimates her MetS posterior probability at 6 months equal to 20.2\% if all the covariates do not change.
The traffic-light algorithm classifies the donor as red (high risk).
A 10\% reduction in BMI (lowering it to 31.7 kg/m$^2$) would yield a substantial decrease in her MetS posterior probability, dropping it to 6.1\%.
While this initial weight loss would not entirely clear the clinical risk, it would shift the donor's profile from the red zone down to the yellow (moderate risk) zone.
This provides a highly motivating feedback loop for the patients, 
showing that even a realistic and achievable reduction in body weight can substantially reduce their predicted risk of MetS.

\section{Discussion and conclusions}
\label{sec:conclusion}

This study focuses on predicting the risk of meeting the diagnostic criteria for MetS at a current pre-donation assessment or any future assessment, by proposing and validating a Bayesian multivariate mixed-effects model that exploits longitudinal clinical and lifestyle data from a predominantly healthy donor population.
Unlike most existing approaches, which rely on cross-sectional data or clinically compromised cohorts, our analysis leverages repeated measurements from blood donors to characterise the temporal evolution and interdependence of metabolic risk factors as donors approach, and in some cases meet, the diagnostic criteria for MetS.
Including subject-specific random effects also allows the model to account for unobserved donor heterogeneity.

By capturing donor-specific heterogeneity and dependence among repeated measurements, the model provides individualised risk estimates that extend beyond population-level averages.
The horseshoe prior regularises the regression coefficients by shrinking weakly supported covariate effects toward zero, thereby highlighting covariates with larger posterior effects while reducing the risk of overfitting.

Moreover, within the Bayesian framework, missing waist-circumference measurements are treated as latent quantities and inferred jointly with the model parameters, while individual MetS risk is obtained from the posterior predictive distribution by assessing the probability that at least three of the five diagnostic criteria are met.

The Bayesian model achieved predictive performance comparable to that of the alternative ML approaches, while providing a balanced classification of MetS cases and non-cases.
Its main advantage, however, lies beyond classification performance: our framework explicitly accounts for the longitudinal and multivariate structure of the data and provides individualised posterior predictive probabilities, allowing predictive uncertainty and individual risk profiles to be incorporated into the screening process.

The posterior predictive probabilities are translated into a three-level traffic-light system representing low, intermediate, and high predicted risk.
Rather than providing a binary classification alone, this stratification may support early identification and clinical prioritisation.
The high sensitivity observed in the validation set (Table~\ref{tab:trafficlight_classification_testset}) indicates that the system can identify most at-risk donors, while the ordered risk categories provide clinicians with an intuitive representation of individual predicted risk.

From a practical perspective, the framework could support preventive screening by leveraging routinely collected data.
Given the limited time available during pre-donation consultations, the traffic-light classification could help AVIS Milan clinicians focus greater attention on donors at higher predicted risk and identify those who may benefit from further evaluation or preventive counselling.
A future implementation of {our framework in} the AVIS Milan clinical decision-support system would allow its practical utility to be evaluated in routine practice.
In addition, regular monitoring in the blood donation setting provides an opportunity to follow changes in metabolic profiles over time and promote timely lifestyle interventions.

Beyond risk stratification, the model identifies the covariates most strongly associated with each MetS component.
Although the prevalence of MetS is low in this selected, generally healthy population, individual metabolic risk factors remain present and may evolve over time.
The longitudinal nature of the AVIS Milan dataset therefore provides a valuable opportunity to investigate changes in metabolic profiles and reinforces the potential role of blood donation centres as settings for preventive screening.

The computational cost of MCMC draws is primarily driven by the large number of donors, although most contribute only a few repeated measurements (typically 3--5, as shown in Supplementary Figure~\ref{fig:hist_donation_count}).
For each donor, the model estimates a $K$-dimensional vector of random intercepts, increasing computational complexity.
Nevertheless, the model can be fitted in about three hours on a laptop with an M4 Pro processor and 24 GB of RAM, suggesting that periodic model fitting may be computationally feasible as new data become available.

Despite its strengths, several limitations warrant acknowledgement.
First, the model's generalizability may be limited by the specific characteristics of the AVIS Milan population, including its demographic composition and eligibility criteria.
External validation in more diverse populations is therefore necessary.
Second, although the traffic-light system facilitates interpretation, the choice of risk thresholds involves a trade-off between sensitivity and specificity and may require recalibration depending on the clinical context.
Moreover, although the model relies on routinely collected clinical and lifestyle variables, it may omit relevant factors such as genetic information, detailed dietary patterns, or socioeconomic determinants, whose inclusion could further improve predictive performance.
Finally, the hypothetical intervention scenarios considered in Section~\ref{sec:colloquium} should not be interpreted causally.
They quantify how posterior predictive risk changes under alternative covariate profiles, conditional on the fitted model, rather than estimating a causal effect of modifying a given risk factor.
Subject to these limitations, our framework provides a flexible and interpretable approach to individualised MetS risk prediction and supports further prospective evaluation in blood-donation settings.

\section*{Acknowledgments}

The authors wish to express their sincere appreciation to AVIS Milan for involving us in their activities, for providing the data, for testing the developed tool, and for the support offered.
Special thanks are extended to the Director Sergio Casartelli for the valuable support throughout the development of this research.
AI was used for English-language editing to improve the grammar and readability of the manuscript.

\section*{Funding}

The first three authors acknowledge the support of MUR, grant Dipartimento di Eccellenza 2023–2027.

\section*{Conflicts of interest}

The authors declare that they have no competing interests.


\bibliographystyle{apalike}
\bibliography{reference}

\vspace{2cm}
\appendix
\begin{center}
    \LARGE \textbf{Appendix}
\end{center}
\vspace{1cm}

\setcounter{section}{0}
\setcounter{equation}{0}
\setcounter{figure}{0}
\setcounter{table}{0}

\renewcommand{\thesection}{\Alph{section}}
\renewcommand{\thesubsection}{\thesection.\arabic{subsection}} 

\renewcommand{\thefigure}{\thesection.\arabic{figure}}
\renewcommand{\thetable}{\thesection.\arabic{table}}
\renewcommand{\theequation}{\thesection.\arabic{equation}}

\makeatletter
    \renewcommand{\p@subfigure}{\thefigure}
    \renewcommand{\theHfigure}{\thesection.\arabic{figure}}
    \renewcommand{\theHtable}{\thesection.\arabic{table}}
    \renewcommand{\theHequation}{\thesection.\arabic{equation}}
\makeatother


\section{Appendix A}
\label{sec:appendixA}

\subsection{Exploratory data analysis}
\label{app:eda}

\begin{figure}[htbp]
\centering
\includegraphics[width=0.7\textwidth]{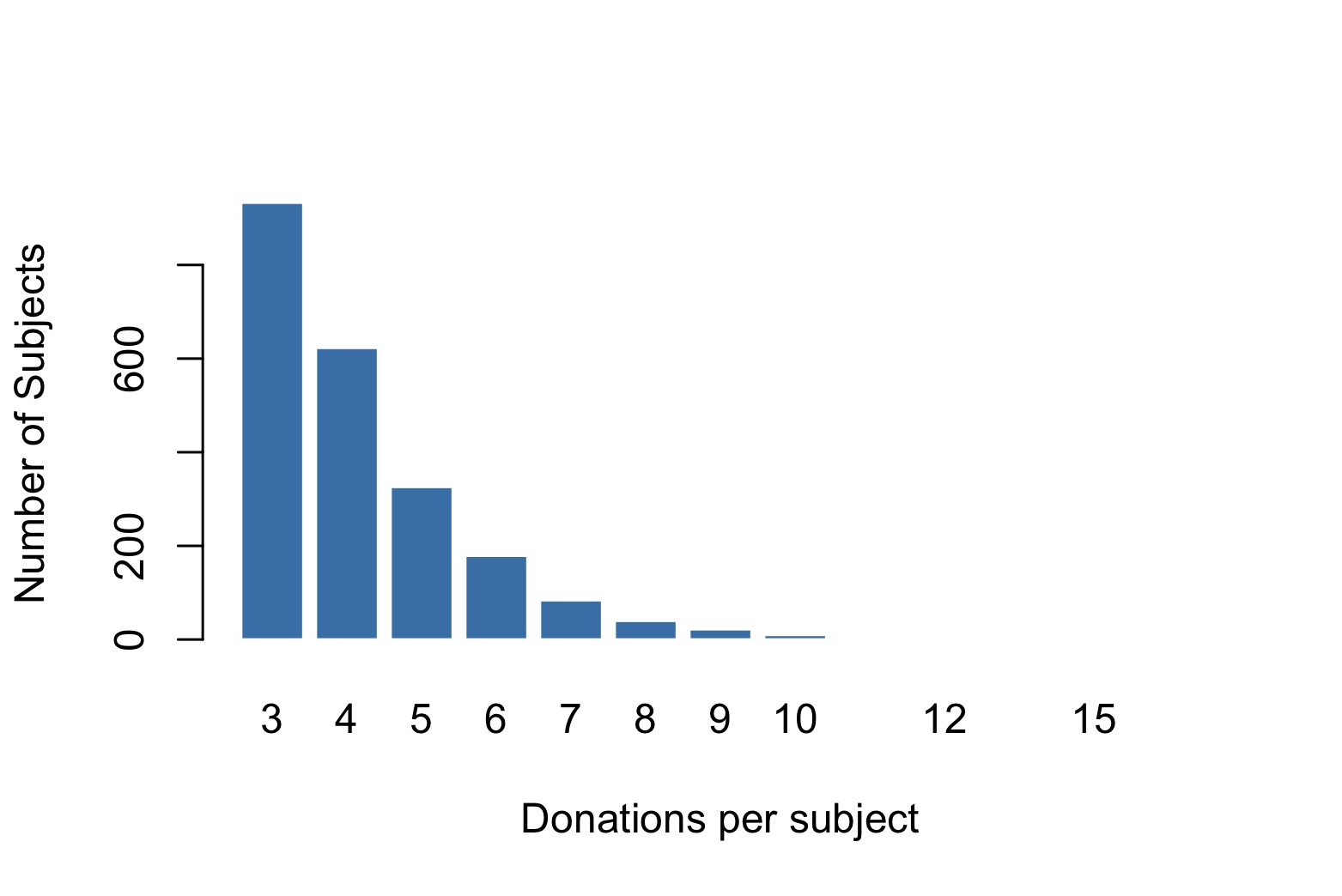}
\caption{Barplot of the donation counts for each donor in the dataset.}
\label{fig:hist_donation_count}
\end{figure}

\begin{figure}[htbp]
\centering
\includegraphics[width=0.9\linewidth]{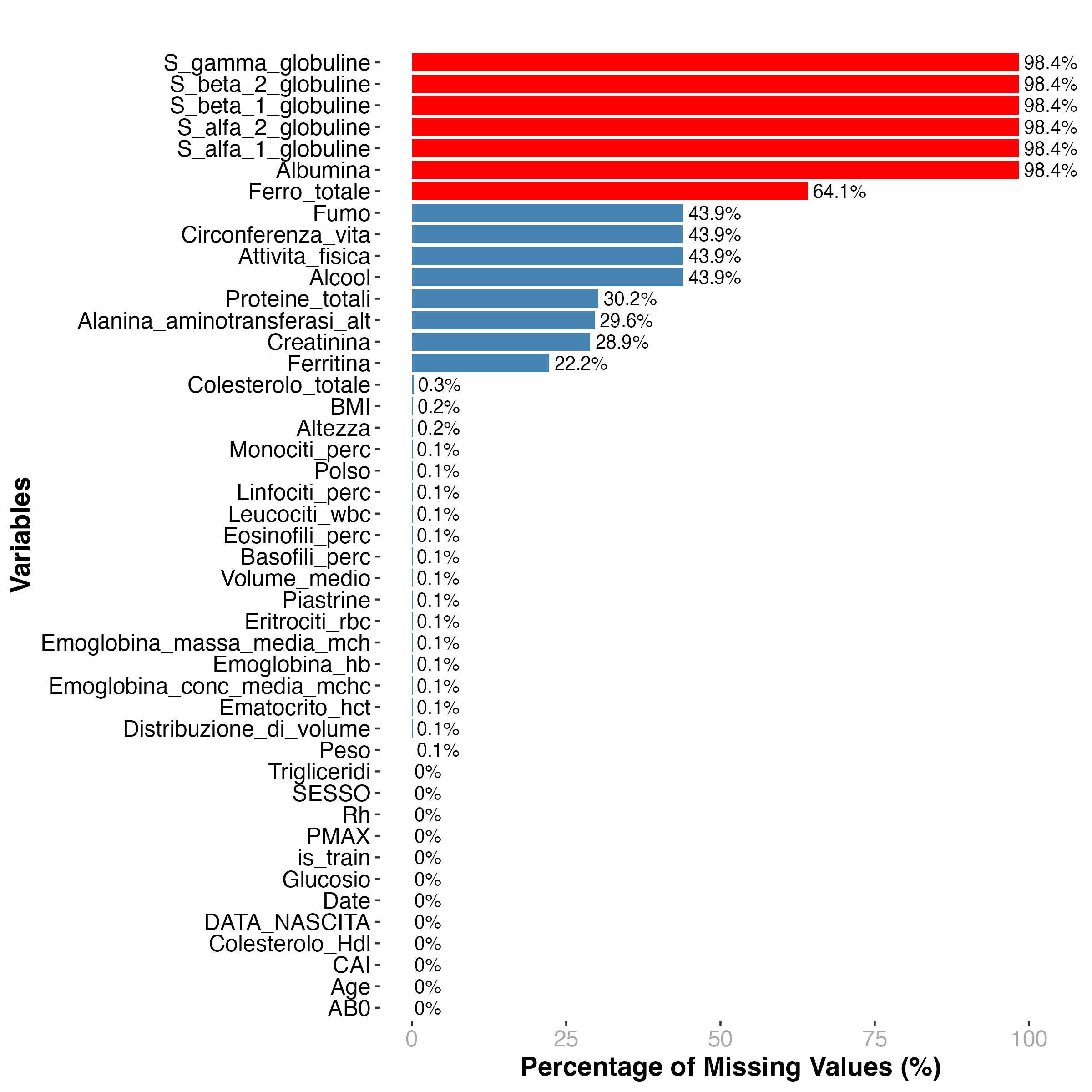}
\caption{Percentage of missing values of each covariate. Red bars represent covariates excluded from the analysis, whereas blue bars denote variables handled through imputation.}
\label{fig:missing_values}
\end{figure}

Figure~\ref{fig:hist_donation_count} reports the barplot of the donation counts for all the donors in the dataset.
The percentage of missing values for each variable in our dataset is shown in Figure~\ref{fig:missing_values}.
Variables with more than 50\% missingness, as \textit{albumin}, \textit{total iron}, \textit{S\_Alpha\_1\_Globulin}, \textit{S\_Alpha\_2\_Globulin}, \textit{S\_Beta\_1\_Globulin}, \textit{S\_Beta\_2\_Globulin}, and \textit{S\_Gamma\_Globulin}, were excluded from analysis to avoid excessive reliance on imputation.
Among the remaining variables, the response variable \textit{waist circumference} exhibits substantial missingness (48\%), primarily because it is measured annually for each donor.
This level of incompleteness motivates a modelling strategy based on data augmentation, in which missing waist measurements are treated as latent variables within the Bayesian framework.
In contrast, the other target variables (\textit{triglycerides}, \textit{PMAX}, \textit{glucose}, \textit{HDL cholesterol}) show very low levels of missingness (below 1\%).
Questionnaire-derived covariates (\textit{alcohol}, \textit{smoking}, \textit{physical activity}) exhibit 46--47\% of missing observations, reflecting the incomplete availability of self-reported information.

\begin{figure}[htbp]
\centering
\includegraphics[width=1\textwidth]{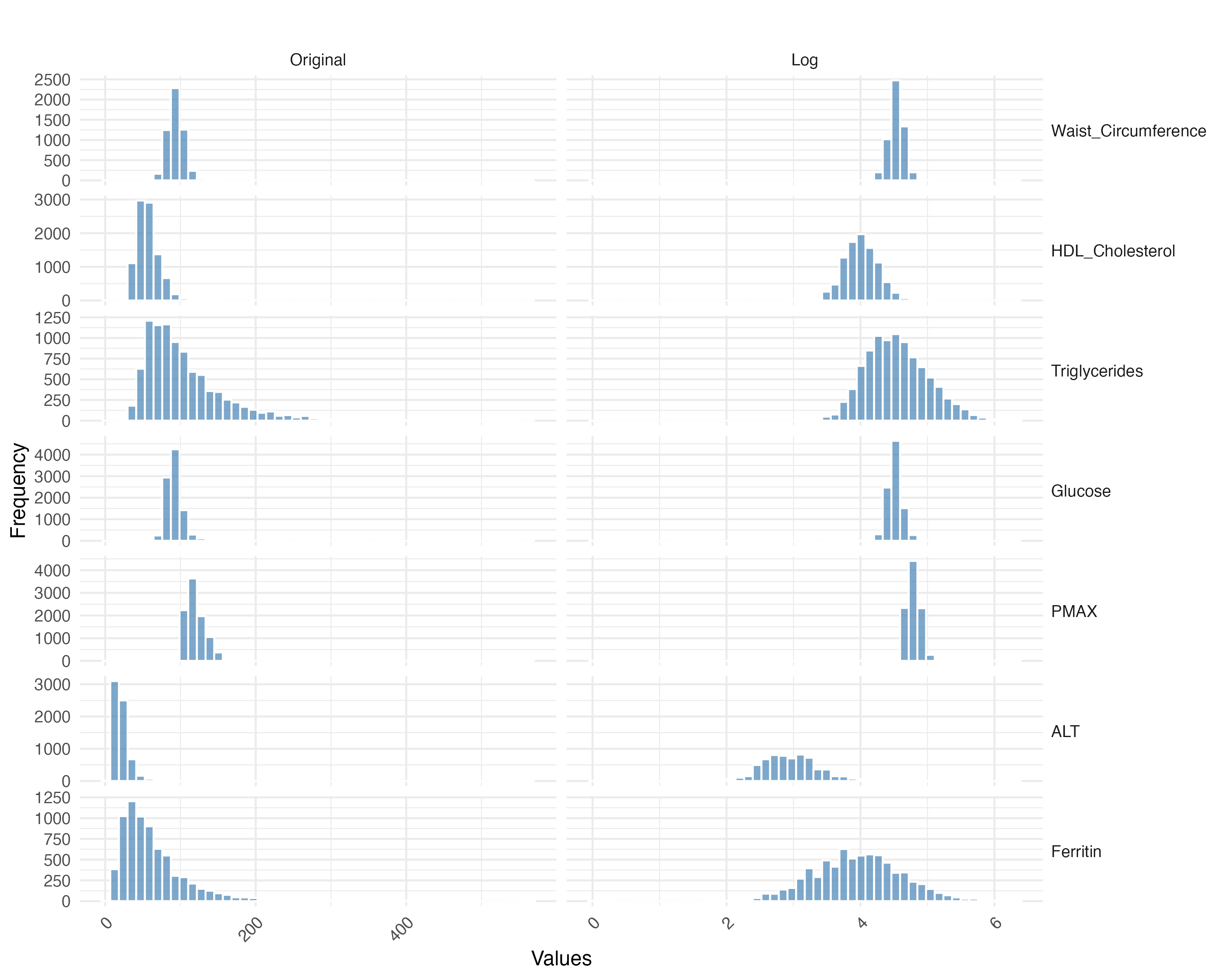}
\caption{Histograms of the five target variables and covariates \textit{ALT} and \textit{ferritin}. For each variable, the histogram is shown on the original scale (left) and on the log scale (right).}
\label{fig:skewed_distributions}
\end{figure}

Examining the shapes of the distributions of the variables in our dataset, we detect right-skewness across the five target variables and covariates \textit{alanine aminotransferase} (\textit{ALT}) and \textit{ferritin}, as shown in Figure~\ref{fig:skewed_distributions}.
Strong right-skewness is evident for \textit{ALT}, \textit{ferritin}, and \textit{triglycerides}, with milder asymmetry observed for \textit{waist circumference}, \textit{glucose}, \textit{PMAX}, and \textit{HDL cholesterol}.
This skewness justifies applying log-transformations to these variables to achieve approximately symmetric distributions.

\begin{table}[htbp]
\centering
\centering
\begin{tabular}[t]{llr}
\hline
\textbf{Variable 1} & \textbf{Variable 2} & \textbf{Pearson $r$} \\
\hline
Haematocrit & Haemoglobin & 0.892 \\
\hline
MCH & Mean\_volume & 0.844 \\
\hline
BMI & Weight & 0.826 \\
\hline
Haematocrit & RBC & 0.799 \\
\hline
Haemoglobin & RBC & 0.697 \\
\hline
MCV & RBC & -0.621 \\
\hline
MCH & MCHC & 0.568 \\
\hline
MCH & RBC & -0.567 \\
\hline
Height & Weight & 0.565 \\
\hline
MCH & Volume\_distribution & -0.459 \\
\hline
MCHC & Volume\_distribution & -0.440 \\
\hline
Age & CAI & -0.395 \\
\hline
Haemoglobin & MCHC & 0.366 \\
\hline
Creatinine & Haemoglobin & 0.339 \\
\hline
Creatinine & Height & 0.333 \\
\hline
Lymphocytes\_pct & WBC & -0.315 \\
\hline
Basophils\_pct & Eosinophils\_pct & 0.309 \\
\hline
Haemoglobin & Weight & 0.306 \\
\hline
Platelets & WBC & 0.295 \\
\hline
Height & Haemoglobin & 0.294 \\
\hline
\end{tabular}
\caption{Top 20 highest absolute Pearson correlations among numerical covariates.}
\label{tab:correlations_1}
\end{table}
We then compute correlations among the numerical covariates in the training set to assess potential multicollinearity.
Table~\ref{tab:correlations_1} shows the top 20 couples of covariates with the highest correlation in absolute value.
Several pairs exhibit high correlations: \textit{haematocrit} with \textit{haemoglobin} ($r=0.892$), \textit{MCH} with \textit{MCV}
($r=0.844$), \textit{BMI} with \textit{weight} ($r=0.826$), and \textit{haematocrit} with \textit{red blood cell count (RBC)} ($r=0.799$).
To improve model interpretability and reduce multicollinearity, we retain \textit{haemoglobin} due to its central physiological importance and removed \textit{haematocrit}, \textit{MCV}, \textit{weight}, and \textit{RBC}.

Additionally, we exclude \textit{total cholesterol} because it has a direct deterministic dependence on \textit{HDL cholesterol} (\textit{Total = HDL + LDL + VLDL}).
Including \textit{total cholesterol} as a predictor would introduce a compositional relationship in which the predictor mathematically contains the outcome itself, leading to a mechanically induced association rather than an independent explanatory effect.
After these adjustments, the highest remaining correlation is $0.568$ between \textit{MCH} and \textit{MCHC}.
Variance Inflation Factors (VIFs) indicate acceptable multicollinearity for all remaining numerical covariates, with all VIFs $< 5$.

\begin{table}[htbp]
\centering
\begin{tabular}[t]{lrrll}
\hline
\textbf{Variable Pair} & \textbf{$\chi^2$} & \textbf{df} & \textbf{p-value} & \textbf{Significant} \\
\hline
\textit{alcohol} vs Physical\_activity & 8.96 & 2 & 1.13e-02 & Yes \\
\hline
Rh vs ABO & 10.78 & 3 & 1.3e-02 & Yes \\
\hline
\textit{smoking} vs \textit{sex} & 5.00 & 1 & 2.53e-02 & Yes \\
\hline
ABO vs \textit{sex} & 3.89 & 3 & 2.74e-01 & No \\
\hline
\textit{alcohol} vs \textit{smoking} & 8.97 & 1 & 2.75e-03 & Yes \\
\hline
\textit{smoking} vs Rh & 0.96 & 1 & 3.26e-01 & No \\
\hline
Physical\_activity vs \textit{sex} & 2.15 & 2 & 3.41e-01 & No \\
\hline
Physical\_activity vs ABO & 13.44 & 6 & 3.66e-02 & Yes \\
\hline
\textit{smoking} vs ABO & 13.45 & 3 & 3.76e-03 & Yes \\
\hline
Physical\_activity vs Rh & 1.25 & 2 & 5.35e-01 & No \\
\hline
\textit{alcohol} vs ABO & 7.59 & 3 & 5.52e-02 & No \\
\hline
Rh vs \textit{sex} & 7.69 & 1 & 5.54e-03 & Yes \\
\hline
\textit{alcohol} vs Rh & 0.09 & 1 & 7.69e-01 & No \\
\hline
Physical\_activity vs \textit{smoking} & 27.89 & 2 & 8.81e-07 & Yes \\
\hline
\textit{alcohol} vs \textit{sex} & 70.91 & 1 & $<$ 2e-16 & Yes \\
\hline
\end{tabular}
\caption{Chi-Squared tests for associations among categorical covariates.}
\label{tab:chi_sq}
\end{table}
We also examine associations among categorical covariates using asymptotic chi-squared tests for all pairwise combinations (Table~\ref{tab:chi_sq}).
The strongest associations (lowest $p$-values) emerge between \textit{alcohol} and \textit{sex} ($\chi^2=70.91$, $p<2\times10^{-16}$), \textit{physical activity} and \textit{smoking} ($\chi^2=27.89$, $p=8.81\times10^{-7}$), and \textit{alcohol} and \textit{smoking} ($\chi^2=8.97$, $p=2.75\times10^{-3}$).
We exclude from the analysis the variable \textit{smoking} and \textit{alcohol} due to the high percentage of missing values and their strong associations with other categorical covariates, such as \textit{alcohol} with \textit{sex} ($\chi^2=70.91$, $p<2\times10^{-16}$), \textit{physical activity} with \textit{smoking} ($\chi^2=27.89$, $p=8.81\times10^{-7}$), and \textit{alcohol} with \textit{smoking} ($\chi^2=8.97$, $p=2.75\times10^{-3}$), which could introduce redundancy and complicate the interpretation of lifestyle effects.

\begin{table}[htbp]
\centering
\begin{tabular}[t]{lrrrrr}
\hline
 & HDL & Waist\_Circumference & \textit{glucose}& \textit{PMAX} & \textit{triglycerides} \\
\hline
ALT & -0.176 & 0.259 & 0.046 & 0.087 & 0.227 \\
\hline
Height & -0.208 & 0.278 & -0.014 & 0.011 & 0.051 \\
\hline
Basophils\_pct & 0.102 & -0.046 & -0.027 & -0.015 & -0.045 \\
\hline
Creatinine & -0.169 & 0.156 & 0.024 & 0.032 & 0.134 \\
\hline
Volume\_distribution & 0.049 & 0.056 & 0.004 & 0.097 & -0.012 \\
\hline
MCHC & -0.139 & 0.043 & -0.051 & 0.006 & 0.073 \\
\hline
Haemoglobin & -0.248 & 0.273 & 0.023 & 0.134 & 0.220 \\
\hline
MCH & 0.091 & -0.073 & -0.004 & -0.052 & -0.023 \\
\hline
Eosinophils\_pct & -0.023 & 0.022 & -0.027 & -0.023 & 0.020 \\
\hline
Ferritin & -0.107 & 0.113 & 0.017 & 0.037 & 0.108 \\
\hline
WBC & -0.140 & 0.169 & 0.035 & 0.071 & 0.207 \\
\hline
Lymphocytes\_pct & 0.045 & -0.050 & -0.036 & -0.099 & -0.053 \\
\hline
Monocytes\_pct & 0.013 & 0.021 & -0.079 & -0.025 & -0.041 \\
\hline
Platelets & 0.047 & 0.020 & 0.016 & 0.025 & 0.039 \\
\hline
Heart\_rate & -0.074 & 0.081 & 0.041 & 0.132 & 0.091 \\
\hline
Total\_proteins & -0.038 & 0.034 & 0.011 & -0.008 & 0.080 \\
\hline
Age & 0.073 & 0.233 & 0.137 & 0.303 & 0.107 \\
\hline
BMI & -0.350 & 0.819 & 0.111 & 0.282 & 0.284 \\
\hline
\end{tabular}
\caption{Pearson correlations: numerical covariates vs MetS targets.}
\label{tab:pearson_target_cov}
\end{table}

We continue the exploratory data analysis by assessing the association between the target variables and the selected covariates to identify potentially uninformative predictors before model specification.
For the numerical covariates, we compute the Pearson correlation coefficient $r$ between each target and each variable (Table~\ref{tab:pearson_target_cov}).
For categorical covariates, we rely on Kruskal–Wallis tests to evaluate differences in target distributions across groups (Table~\ref{tab:kw_target_cov}).
Based on these results, we exclude the categorical covariates \textit{Rh} and \textit{ABO}.
Although some associations with the targets are statistically significant, they are generally weak and accompanied by associations with other categorical covariates (see Table~\ref{tab:chi_sq}), suggesting limited additional explanatory value.
After this screening step, we include the 18 numerical covariates reported in Table~\ref{tab:pearson_target_cov} in the model.
Among categorical variables, we include \textit{sex}, given its consistently strong associations with all target variables (Table~\ref{tab:kw_target_cov}), and \textit{physical activity}, due to its relevance as a key lifestyle factor and its systematic association with metabolic outcomes.

\begin{figure}[htbp]
\centering
\includegraphics[angle=90, width=\textheight, height=1.3\textwidth, keepaspectratio]{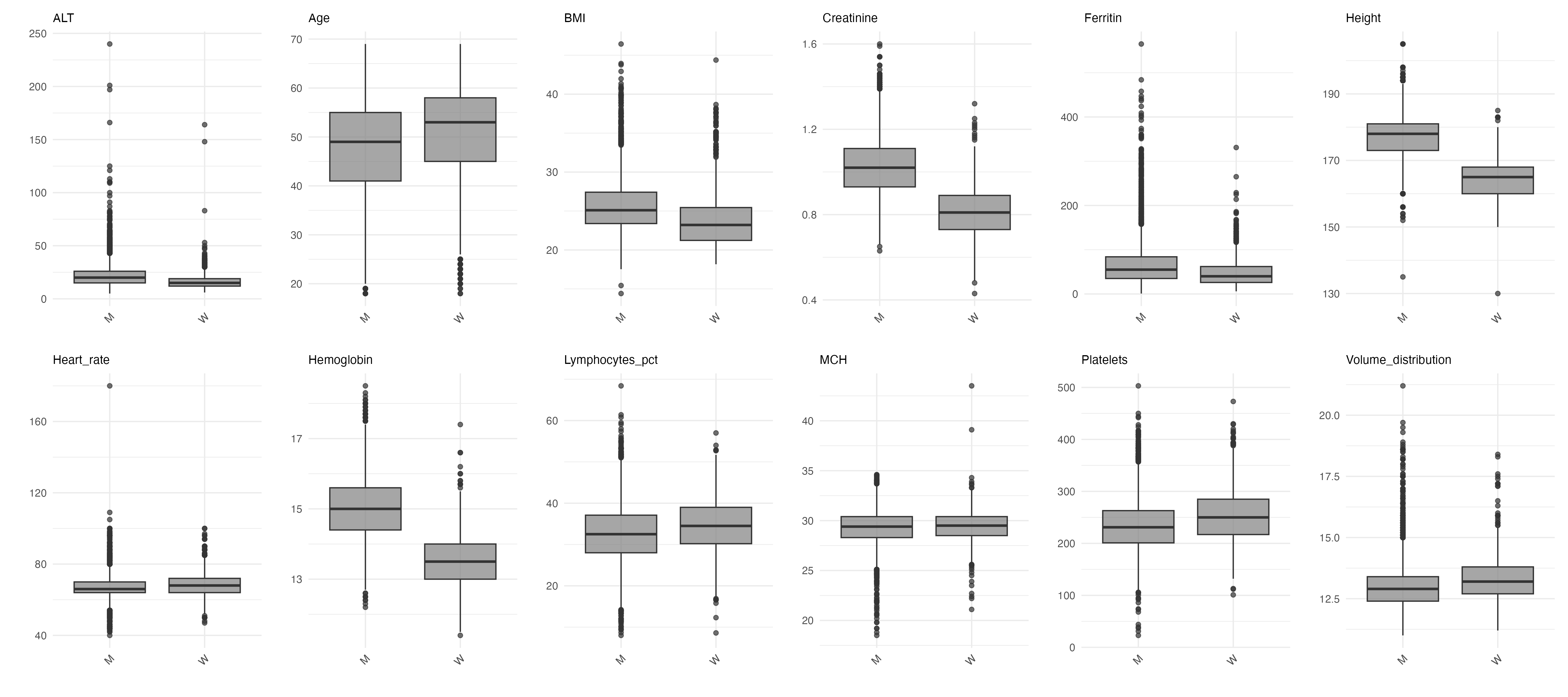}
\caption{Boxplots of numerical covariates grouped by \textit{sex}.}
\label{fig:boxplots}
\end{figure}

\begin{table}[htbp]
\centering
\begin{tabular}[t]{llllll}
\hline
Covariate & HDL & Waist\_Circumference & \textit{glucose}& \textit{PMAX} & \textit{triglycerides} \\
\hline
Physical\_activity & $<$2e-16 & $<$2e-16 & 1.6e-06 & 3.23e-07 & $<$2e-16 \\
\hline
Rh & 4.63e-02 & 1.95e-01 & 1.26e-01 & 9.44e-01 & 3.31e-02 \\
\hline
ABO & 4.91e-04 & 2.28e-03 & 4.08e-02 & 4.54e-02 & 1.69e-04 \\
\hline
\textit{sex} & $<$2e-16 & $<$2e-16 & 8.87e-07 & $<$2e-16 & $<$2e-16 \\
\hline
\end{tabular}
\caption{Kruskal-Wallis p-values: categorical covariates vs METs targets.}
\label{tab:kw_target_cov}
\end{table}
Finally, we explore the associations between the remaining categorical and numerical covariates using Kruskal-Wallis tests (Table~\ref{tab:kw_covariates_only}) and boxplots (Figure~\ref{fig:boxplots}).
These analyses reveal clear distributional shifts across \textit{sex} for several numerical covariates and target variates.
To account for such heterogeneity, we include interaction terms between \textit{sex} and the numerical covariates highlighted in Figure~\ref{fig:boxplots} in the final model specification.
Additionally, we introduce the interaction between \textit{sex} and \textit{physical activity} to investigate 
potential sex-specific lifestyle effects on MetS risk for men and women.

\begin{table}[htbp]
\centering
\begin{tabular}[t]{lll}
\hline
Numerical covariate & \textit{physical activity} & \textit{sex} \\
\hline
ALT & 4.45e-04 & $<$2e-16 \\
\hline
Height & 2.4e-01 & $<$2e-16 \\
\hline
Basophils\_pct & 1.03e-05 & 3.96e-02 \\
\hline
Creatinine & 2.39e-01 & $<$2e-16 \\
\hline
Volume\_distribution & 4.81e-01 & $<$2e-16 \\
\hline
MCHC & 2.25e-01 & $<$2e-16 \\
\hline
Haemoglobin & 6.56e-03 & $<$2e-16 \\
\hline
MCH & 4.22e-05 & 1.29e-01 \\
\hline
Eosinophils\_pct & 4.48e-02 & $<$2e-16 \\
\hline
Ferritin & 3.56e-01 & $<$2e-16 \\
\hline
WBC & 2.76e-13 & 2.37e-01 \\
\hline
Lymphocytes\_pct & 1.74e-02 & $<$2e-16 \\
\hline
Monocytes\_pct & 7.76e-03 & $<$2e-16 \\
\hline
Platelets & 3.6e-01 & $<$2e-16 \\
\hline
Heart\_rate & 1.49e-10 & 2.4e-07 \\
\hline
Total\_proteins & 1.88e-02 & 7.36e-15 \\
\hline
Age & 2.34e-03 & $<$2e-16 \\
\hline
BMI & $<$2e-16 & $<$2e-16 \\
\hline
\end{tabular}
\caption{Kruskal-Wallis p-values: continuous responses vs categorical covariates.}
\label{tab:kw_covariates_only}
\end{table}

\subsection{Missing values in the covariates}
\label{app:missing_cov}

The dataset contains missing values in several covariates (see Figure~\ref{fig:missing_values}).
These are imputed using multiple imputation by chained equations (MICE)~\citep{van2011mice} under a Fully Conditional Specification (FCS) framework~\citep{van2012flexible}, with two-level predictive mean matching (2l.pmm) implemented in the add-on package \texttt{miceadds}~\citep{robitzsch2021miceadds}.
Donor ID is specified as the level-2 clustering variable to account for the longitudinal structure of repeated measurements.

Variables excluded from our Bayesian model (as explained above), such as \textit{ABO} and \textit{Rh}, are retained as auxiliary predictors in the imputation procedure to improve the estimation of missing-value distributions.
The target variables were intentionally excluded as predictors in the imputation models to avoid inducing spurious relationships between the outcomes and the covariates.
The MICE algorithm is run for 10 iterations.

\section{Appendix B}
\label{sec:appendixB}

\setcounter{figure}{0}
\setcounter{table}{0}
\renewcommand{\thefigure}{B.\arabic{figure}}
\renewcommand{\thetable}{B.\arabic{table}}

This appendix summarises the model-building strategy that led to the final multivariate Bayesian model.
We have progressively investigated increasingly complex dependence structures to determine whether explicit temporal autoregression improves predictive performance beyond the multivariate mixed-effects formulation adopted in the main text.

\subsection{Univariate t-CAR models}
\label{app:univariate_temporalDependence}

We have investigated potential temporal dependencies by fitting five independent univariate mixed-effects models.
At the initial visit ($j=1$), for the target variable $k$ the outcome depends exclusively on the covariates and the donor-specific random intercept:
$$Y_{i1}^{(k)} \sim \mathcal{N}(\mu_{i1}^{(k)}, \sigma_k^2)$$
$$\mu_{i1}^{(k)} = x_{i1}^\prime \bm{\beta}^{(k)} + b_i^{(k)}$$ 
For any subsequent visit ($j > 1$), the linear predictor is augmented with an exponentially decaying autoregressive term, incorporating the outcome from the previous visit $Y_{i,j-1}^{(k)}$ and the elapsed time $\Delta t_{ij}$ (measured in years):
$$Y_{ij}^{(k)} \sim \mathcal{N}(\mu_{ij}^{(k)}, \sigma_k^2)$$
$$\mu_{ij}^{(k)} = \bm x_{ij}^\prime \bm \beta^{(k)} + b_i^{(k)} + \exp(-\phi^{(k)} \Delta t_{ij}) Y_{i,j-1}^{(k)}$$

We used a $\text{Gamma}(2, 0.2)$ marginal prior for target-specific decay rate parameters and the same marginal priors as the original model for the other parameters

Table~\ref{tab:phi_univariate_summary} reports the posterior summaries for the decay parameters estimated by these independent models.

\begin{table}[htbp]
\centering
\begin{tabular}{lcccc}
\toprule
\textbf{Target Variable} & \textbf{Median} & \textbf{Mean} & \textbf{95\% CI Lower} & \textbf{95\% CI Upper} \\
\midrule
\textit{PMAX} $\phi^{(1)}$ & 15.37 & 16.62 & 10.88 & 30.08 \\
\textit{HDL cholesterol} $\phi^{(2)}$& 21.44 & 23.04 & 14.89 & 34.33 \\
\textit{glucose}$\phi^{(3)}$ & 19.44 & 20.86 & 12.72 & 37.53 \\
\textit{triglycerides} $\phi^{(4)}$ & 21.80 & 23.11 & 14.11 & 39.79 \\
\textit{waist circumference }$\phi^{(5)}$ & 25.53 & 26.66 & 14.91 & 45.84 \\
\bottomrule
\end{tabular}
\caption{Posterior summaries of the decay rate parameter $\phi$ from the independent univariate t-CAR models.}
\label{tab:phi_univariate_summary}
\end{table}
Posterior inference is highly consistent with the multivariate model estimates.
The estimated decay rates remain notably high across all five independently fitted models.
Applying the same conservative assessment on the lower bounds of the 95\% credible intervals, the slowest decay is observed for systolic blood pressure \textit{PMAX} ($\phi^{(1)} \approx 10.88$).
Even in this worst-case scenario, the residual autocorrelation drops to $\approx 0.066$ after three months ($\Delta t = 0.25$) and is practically zero ($\approx 0.004$) after six months ($\Delta t = 0.5$).
For the highest lower bound (\textit{waist circumference}, $\phi^{(5)} \approx 14.91$), the autocorrelation falls to $\approx 0.024$ at three months.

Considering the average intervals between donations, these univariate analyses independently corroborate that temporal autocorrelation vanishes well before the subsequent visit.
This provides further solid justification for omitting the autoregressive component in the final model specifications.

\subsection{Model I: multivariate t-CAR model with time-invariant fixed effects}
\label{app:temporaDependence}

To further investigate potential temporal dependencies, we have extended our Bayesian multivariate GLMM \eqref{eq:ver}–\eqref{eq:prior_randomeff_param2} in the manuscript to explicitly account for the unequally spaced longitudinal nature of the donations.
The covariates, evaluated at the same time as the target variables, enter the model through linear predictors that do not change over time.
Specifically, we incorporated a Continuous Autoregressive (t-CAR) structure \citep{diggle1988approach}.
In particular, for each donor~$i$ the linear predictor of the log-transformed and standardised metabolic target $Y_{ij}^{(k)}$ at any subsequent visit $j > 1$ is updated to include an autoregressive term of $Y_{i \ j-1}^{(k)}$
that decays exponentially with time $\Delta t_{i j}$, measured in years, elapsed since the previous visit.
Specifically, we assume the likelihood \eqref{eq:ver} in the manuscript with 
\begin{equation*}
\mu_{i j}^{(k)} = \mathbf{x}_{i j}^\prime \bm{\beta}^{(k)} + b_i^{(k)} + \exp(-\phi^{(k)} \Delta t_{i j}) Y_{i \ j-1}^{(k)}, \, j=2,\ldots,n_i.
\end{equation*}
For the first visit ($j=1$), the mean vector is defined exclusively by the time-invariant covariates and the donor-specific random intercepts $b_i \in \mathbb{R}^K$.
Here, $\phi^{(k)}\geq 0$ represents the target-specific decay rate of the temporal autocorrelation.
To ensure positivity, we assume a weakly informative Gamma prior to these parameters, specifically $\phi^{(k)}\iid \text{Gamma}(2, 0.2)$, $k=1,2,\ldots,5$.
The rest of the marginal priors are fixed as in Section~\ref{sec:mult_GLMM} of the manuscript.
 
Table~\ref{tab:phi_summary} reports the posterior summaries for the decay rate parameters $\phi^{(1)}, \dots, \phi^{(5)}$.

\begin{table}[htbp]
\centering
\begin{tabular}{lcccc}
\toprule
\textbf{Target Variable} & \textbf{Median} & \textbf{Mean} & \textbf{95\% CI Lower} & \textbf{95\% CI Upper} \\
\midrule
\textit{PMAX} $\phi^{(1)}$ & 14.97 & 16.15 & 10.60 & 28.91 \\
\textit{HDL cholesterol} $\phi^{(2)}$ & 18.60 & 19.95 & 13.08 & 34.68 \\
\textit{glucose}$\phi^{(3)}$ & 19.03 & 20.48 & 12.55 & 36.69 \\
\textit{triglycerides} $\phi^{(4)}$ & 23.95 & 25.32 & 15.95 & 42.36 \\
\textit{waist circumference }$\phi^{(5)}$ & 24.36 & 25.94 & 14.27 & 45.33 \\
\bottomrule
\end{tabular}
\caption{Posterior summaries of the decay rate parameters $\phi^{(1)}, \dots, \phi^{(5)}$ from the multivariate t-CAR model.}
\label{tab:phi_summary}
\end{table}

The posterior estimates of $\phi^{(1)}, \ldots, \phi^{(5)}$ are notably high across all five target variables.
In the context of a t-CAR process, such values 
indicate that the correlation between consecutive measurements drops toward zero extremely quickly.
To provide the most conservative assessment, we can evaluate the residual autocorrelation $\exp(-\phi^{(k)} \Delta t_{i j})$ using the lower bounds of the 95\% credible intervals (i.e., the slowest estimated decay rates).
For systolic blood pressure, which exhibits the lowest overall bound ($\phi \approx 10.61$), autocorrelation after just three months ($\Delta t = 0.25$) is only $\approx 0.071$ (roughly 7\%).
After six months ($\Delta t = 0.5$), it drops to negligible $\approx 0.005$.
Instead, for \textit{triglycerides}, which presents the highest lower bound ($\phi \approx 15.96$), the autocorrelation plummets to $\approx 0.018$ after three months and is virtually zero ($\approx 0.0003$) after six months.

Given that the sample average interval between consecutive visits is 291 days for men and 339 days for women, the resulting autocorrelation factor becomes negligibly small for all practical purposes, even under the most conservative posterior bounds.
These results confirm that temporal autocorrelation is entirely negligible once the relevant covariates and donor-specific random intercepts are accounted for.

\subsection{Sequential autoregressive models with dynamic time-varying effects}
\label{app:autoregressive_models}

To explicitly investigate the temporal effects of covariates from previous visits, we consider a sequential autoregressive model formulation.
We assume, as in the previous section, that $\bm Y_{ij}$ follows the multivariate normal distribution \eqref{eq:ver} (in the manuscript).
To properly specify the linear predictor, we partition the covariate vector $\bm{x}_{ij} \in \mathbb{R}^P$ into a set of time-invariant predictors $\bm x_{ij}^{(const)}$ and a set of time-varying predictors $\bm x_{ij}^{(tv)}$.
Let $\bm \beta^{(const)}$ and $\bm \beta^{(tv)}$ denote the corresponding matrices of regression coefficients.
For the first visit ($j=1$), the mean vector is defined exclusively by the time-invariant covariates and the donor-specific random intercepts ${\bm b}_i \in \mathbb{R}^K$, i.e., $\bm \mu_{i1} = (\bm \beta^{(const)})^\prime \bm x_{i1}^{(const)}+ \bm b_i$.

For subsequent visits ($j>1$), the mean vector incorporates an autoregressive component based on the biomarker values observed at the previous visit, ${\bm Y}_{i \, j-1}$.
This contribution decays exponentially with the elapsed time $\Delta t_{ij}$ (measured in years) since the previous donation, so the strength of temporal dependence decreases as the interval between consecutive visits increases.
Letting $\alpha \in \mathbb{R}_+^K$ denote the vector of target-specific decay rates, the autoregressive contribution is expressed through the Hadamard product $\exp(-\alpha \Delta t_{ij}) \odot Y_{i,j-1}$.
In addition to this autoregressive component, we examine three model variants that differ in how they dynamically represent the effect of the time-varying covariates.

\paragraph{Model II: static time-varying effects.}

In the first specification, the effect of the time-varying covariates at visit $j$ is assumed to be independent of the time gap $\Delta t_{ij}$.
The linear predictor is formulated as:
\begin{equation*}
 \bm \mu_{ij} = ( \bm \beta^{(const)})^\prime \bm x_{ij}^{(const)} + (\bm\beta^{(tv)}) ^\prime \bm x_{i\ j-1}^{(tv)} + \bm b_i + \exp(- \bm \alpha \Delta t_{ij}) \odot \bm Y_{i,j-1}
\end{equation*}

\paragraph{Model III: exponentially decaying time-varying effects.}

The second model assumes that the informational weight of the time-varying covariates decays exponentially as the time interval $\Delta t_{ij}$ increases:
\begin{equation*}
 \bm \mu_{ij} = (\bm \beta^{(const)})^\prime \bm x_{ij}^{(const)}+ \exp(-\Delta t_{ij}) ( \bm \beta^{(tv)})^\prime\bm x_{i \ j-1}^{(tv)} + \bm b_i + \exp(-\bm \alpha \Delta t_{ij}) \odot \bm Y_{i,j-1}
\end{equation*}

\paragraph{Model IV: linearly scaled time-varying effects.}

The third specification investigates whether the magnitude of the effect of time-varying covariates scales linearly with the elapsed time:
\begin{equation*}
 \bm \mu_{ij} = (\bm \beta^{(const)} )^\prime \bm x_{ij}^{(const)} + \Delta t_{ij} (\bm \beta^{(tv)} )^\prime \bm x_{i\ j-1}^{(tv)} +\bm b_i + \exp(-\bm \alpha \Delta t_{ij}) \odot \bm Y_{i,j-1}
\end{equation*}

In all three configurations, the regression coefficient matrices $\bm \beta^{(const)}$ and $\bm \beta^{(tv)}$ were assigned the same regularizing Horseshoe prior described in Section~\ref{sec:mult_GLMM} of the manuscript, 
in order to handle the high dimensionality of the covariate space and induce sparsity.

\subsection{Predictive model comparison}
\label{app:model_selection}
To implement cross-validation, we divided the data into training, validation, and test sets.
The test set includes independent donors, i.e., donors not included in the other sets.
Specifically, we randomly selected 51 of the 2,228 donors using stratified sampling to preserve the original male-to-female ratio.
We used the remaining 2,177 donors to construct the training set and the validation set.
For each donor, we reserved the last available donation for the validation set, while including all preceding donations in the training set, yielding 9,125 training observations.
Among these 2,177 donors, only 689 had complete information on \textit{waist circumference} at their last donation.

First of all, we compare the predictive performance on the test set of the different model  specifications using LOOIC and WAIC for the model in Section~\ref{sec:mult_GLMM} in the manuscript, the univariate formulation in Supplementary Section~\ref{app:univariate_temporalDependence} and the 
multivariate t-CAR model with time-invariant fixed effects (Model I) in Supplementary Section~\ref{app:temporaDependence}.
For the independent univariate framework, the total information criteria were calculated as the sum of the individual models' values, with the aggregated standard errors (SE) computed as the square root of the sum of the squared individual standard errors.
Lower values of WAIC and LOOIC indicate better model fit.
Table~\ref{tab:ic_comparison} summarises the predictive performance metrics.

\begin{table}[htbp]
\centering
\begin{tabular}{lcccc}
\toprule
\textbf{Model Specification} & \textbf{WAIC} & \textbf{(SE)} & \textbf{LOOIC} & \textbf{(SE)} \\
\midrule
Independent Univariate Models (Sum) & 137,904.7 & (1,745.6) & 139,309.5 & (1,484.8) \\
Multivariate t-CAR (Model I) & 71,838.1 & (462.8) & 74,862.9 & (459.9) \\
\textbf{Multivariate GLMM (Section \ref{sec:mult_GLMM})} & \textbf{71,773.2} & \textbf{(464.2)} & \textbf{74,769.6} & \textbf{(461.9)} \\
\bottomrule
\end{tabular}
\caption{Comparison of predictive performance metrics on the test set across the modeling frameworks. The LOOIC is approximated using the Pareto-smoothed importance sampling method.}
\label{tab:ic_comparison}
\end{table}

Our Bayesian multivariate model described by \eqref{eq:ver}–\eqref{eq:prior_randomeff_param2} in Section~\ref{sec:mult_GLMM} of the manuscript yields a drastically lower LOOIC ($\approx 74,770$) compared to the aggregated independent univariate models ($\approx 139,310$).
This massive improvement in predictive accuracy is primarily driven by the joint model's ability to borrow strength across the correlated metabolic targets.
Specifically, the multivariate structure drastically reduces the uncertainty in imputing the highly prevalent missing values for \textit{waist circumference} by dynamically leveraging the concurrent observed levels of \textit{glucose}, \textit{triglycerides}, blood pressure, and \textit{HDL cholesterol}, through the estimates of the residual and random-effect covariance matrices, $\bm{\Sigma}_t$ and $\bm{\Sigma}_b$.

Furthermore, when comparing the multivariate formulations, the inclusion of the t-CAR component (Supplementary Section~\ref{app:temporaDependence}) does not yield any predictive gain.
The LOOIC for the t-CAR model ($\approx 74,863$) is slightly higher than that of the final simpler specification in \eqref{eq:ver}–\eqref{eq:prior_randomeff_param2} ($\approx 74,770$).
The difference in LOOIC between these two models ($\approx 93$) is substantially smaller than their respective standard errors ($\approx 460$), indicating that the models are practically indistinguishable in terms of expected predictive accuracy.
This minor increase reflects a penalisation for including unnecessary autoregressive parameters without any corresponding improvement in fit.
The WAIC values perfectly mirror this behaviour.
These predictive performance metrics, coupled with the rapid decay of the temporal correlation parameters $\phi^{(k)}$ discussed in Supplementary Sections~\ref{app:temporaDependence} and \ref{app:univariate_temporalDependence}, confirm that the multivariate mixed-effects likelihood \eqref{eq:ver}-\eqref{eq:lin_pred} in the manuscript represents a flexible and parsimonious model for this dataset.

To provide a definitive assessment of the modelling strategies investigated in this study, we compare the Bayesian multivariate GLMM (described in Section \ref{sec:mult_GLMM} in the manuscript) against the continuous autoregressive specification (t-CAR, Appendix \ref{app:temporaDependence}) and the three sequential autoregressive variations equipped with dynamic time-varying effects (Appendix \ref{app:autoregressive_models}).
The models are evaluated based on their goodness-of-fit and parsimony using the Watanabe-Akaike Information Criterion (WAIC) and the Leave-One-Out Information Criterion (LOOIC) computed on the training set.
Furthermore, we assess the predictive accuracy on a test set of donors (i.e. not included in the original training set).
For this aim we use continuous error metrics, namely the Root Mean Square Error (RMSE) and Mean Absolute Error (MAE).
Finally, the discriminative power of the models for the clinical binary classification of Metabolic Syndrome at the last visit in the validation set is summarised by the Area Under the ROC Curve (AUC).
For the univariate models, given their significantly higher WAIC values and non-negligible computational time, we did not compute any prediction metrics such as RMSE or MAE or AUC in the validation set.

Table~\ref{tab:full_model_comparison} reports the performance metrics for all five multivariate formulations.
\begin{table}[htbp]
\centering
\begin{tabular}{lccccc}
\toprule
& \multicolumn{2}{c}{\textbf{Training set}} & \multicolumn{2}{c}{\textbf{Test set}} & \multicolumn{1}{c}{\textbf{Validation set}} \\
\cmidrule(lr){2-3} \cmidrule(lr){4-5} \cmidrule(lr){6-6}
\textbf{Model Specification} & \textbf{WAIC} & \textbf{LOOIC} & \textbf{RMSE} & \textbf{MAE} & \textbf{AUC} \\
\midrule
\textbf{Multivariate GLMM \eqref{eq:ver}-\eqref{eq:prior_randomeff_param2}} & 71921.5 & 74902.8 & 0.849 & 0.659 & 0.942 \\
Multivariate t-CAR (Model I) & 71985.8 & 74944.0 & 0.849 & 0.658 & 0.941 \\
Sequential AR (Model II: Static TVC) & 74773.6 & 77674.7 & 1.006 & 0.785 & 0.945 \\
Sequential AR (Model III: Exp. Decay) & 74747.1 & 77687.8 & 1.014 & 0.790 & 0.943 \\
Sequential AR (Model IV: Linear Scale) & 74813.4 & 77748.9 & 1.008 & 0.785 & 0.945 \\
\bottomrule
\end{tabular}
\caption{Comparison of information criteria and predictive performance metrics across all evaluated multivariate models. Metrics are categorised by their evaluation data subset: WAIC and LOOIC on the training set, RMSE and MAE on the test set, and AUC on the validation set.}
\label{tab:full_model_comparison}
\end{table}

While the inclusion of sequential autoregressive components and dynamic time-varying covariates captures fine-grained temporal mechanics, these highly parameterized structures do not yield substantial improvements in accuracy (RMSE, MAE) on the test set or clinical discrimination (AUC) compared to the multivariate GLMM \eqref{eq:ver}-\eqref{eq:prior_randomeff_param2} in the manuscript.
Moreover, the multivariate GLMM and the t-CAR formulation (Model I) show the shortest runtimes, completing $1,000$ iterations across each of four parallel chains in about 4 hours on a MacBook Pro equipped with an M4 Pro chip.
In contrast, estimating the three Sequential AR specifications (Models II, III, and IV) requires roughly 16 hours to perform the same number of iterations.
Given the negligible differences in predictive performance and the higher computational cost associated with the dynamic components, the multivariate GLMM \eqref{eq:ver}-\eqref{eq:prior_randomeff_param2} (in the manuscript) provides the best balance between predictive performance, parsimony, computational efficiency, and model interpretability.

\section{Appendix C}
\label{sec:appendixC}

\setcounter{figure}{0}
\setcounter{table}{0}
\renewcommand{\thefigure}{C.\arabic{figure}}
\renewcommand{\thetable}{C.\arabic{table}}

\subsection{Simulation design}
\label{sec:sim_design}

To further evaluate the predictive performance of our multivariate Bayesian framework, we have designed a simulation study that emulates the longitudinal dynamics of the AVIS Milan blood donor cohort while introducing temporal dependencies and heavy-tailed errors that are not explicitly accommodated by the fitted model.

The simulation study design assumes three covariates measured at different visits, \textit{BMI}, \textit{age}, and \textit{sex}, which are central determinants of the five MetS biomarker outcomes in the real-case application.
Age and BMI are treated as time-varying covariates, whereas sex is time-invariant.
Age allows the model to account for calendar time across donations.
We have simulated a cohort of $N=50$ donors (43 males, 7 females), each observed at five consecutive visits equally spaced over time.
To mimic the AVIS Milan longitudinal structure in which the average gap time is approximately 10 months, visits have been assumed to occur every 12 months for all donors.

To preserve biological realism, the continuous covariates \textit{BMI} and \textit{age} at the beginning of the study have been sampled directly from the sex-specific empirical distributions in the AVIS Milan dataset.
For donors aged~$\geq 50$ years, the longitudinal trajectory of \textit{BMI} 
is allowed to increase over the visits according to:
$$ \text{BMI}_{ij} = \begin{cases} 
1.05 \cdot \text{BMI}_{i\,j-1} + \epsilon^{\text{BMI}}_{ij} & \text{if } \text{age}_{ij} \geq 50 \\
\text{BMI}_{i\,j-1} + \epsilon^{\text{BMI}}_{ij} & \text{otherwise}
\end{cases} \, , $$
where $\epsilon^{\text{BMI}}_{ij} \iid \mathcal{N}(0, 0.36)$ represent the inter-visit measurement noises.
This model reflects the physiological weight gain often observed in older populations.

The five MetS biomarkers, denoted by the vector $\mathbf{Y}_{ij}$, have been simulated on the same log-transformed and standardised scale used for model fitting.
Their conditional mean is determined by the simulated covariates through the linear predictor $\boldsymbol{\beta}^{\prime}\mathbf{x}_{ij}$, while an additional first-order autoregressive AR(1) component, with a persistence parameter $\rho = 0.5$, is introduced to induce temporal dependence between consecutive visits.
To explicitly test the model's capacity to handle temporal autocorrelation, we have generated the latent outcomes via a first-order autoregressive AR(1) process, and heavy tails:
$$\mathbf{Y}_{ij} = \boldsymbol{\beta}^\prime \mathbf{x}_{i j} + \rho \mathbf{Y}_{i\,j-1} + \sigma\cdot \boldsymbol{\epsilon}_{ij},$$
where $\mathbf{x}_{i j}$ denotes the vector of simulated covariates and $\rho=0.5$ controls the temporal dependence between consecutive visits.
This relatively strong temporal persistence is deliberately introduced to challenge the model under substantial temporal dependence.
The five heavy-tailed innovations in $\boldsymbol{\epsilon}_{ij}$ are generated independently from Student's $t$ distributions with $\nu=4$ degrees of freedom, rather than from the Gaussian distribution assumed by the fitted model.
The scale parameter $\sigma$ is set to 3 to generate appreciable variability in the simulated biomarkers while retaining separation between lower- and higher-risk profiles.
Thus, both the autoregressive component and the heavy-tailed innovations represent deliberate departures from the fitted model's assumptions, allowing us to assess its predictive robustness under model misspecification.

\subsection{Predictive evaluation}
\label{sec:sim_prediction}

For each donor, simulated observations from the first four visits were used for model fitting, while the fifth visit is held out and used as the ground truth for evaluating one-step-ahead predictive performance.
We then applied the traffic-light warning system described in Section~\ref{sec:colloquium} of the manuscript to the posterior predictions for the fifth visit.

We computed the classification boundaries for the simulated dataset using the {same approach as in Section~\ref{sec:colloquium} of the manuscript, deriving the optimal threshold for these data: $0.11$ between yellow and red classes, and $0.04$ between green and yellow classes}.
Table~\ref {tab:sim_results} summarises the predictive performance of the warning system on the simulated fifth-visit observations.
\begin{table}[htbp]
\centering
\begin{tabular}{lcc}
\hline
\textbf{Traffic-light classification} & True No & True Yes \\
\hline
Green & 28 & 2 \\
Yellow & 9 & 2 \\
Red & 5 & 4 \\
\hline
\end{tabular}
\caption{Confusion matrix: traffic-light classification on (simulated) data from the 5th visit.}
\label{tab:sim_results}
\end{table}
When Yellow and Red classifications are jointly regarded as warning signals, the model identifies 6 of the 8 donors who develop MetS at the fifth visit, corresponding to a sensitivity of 75\%.
Among the 42 donors who remain MetS-negative, 28 are assigned to the Green category, corresponding to a specificity of 66.7\%.
Thus, despite the model misspecification given by the temporal dependence and heavy-tailed errors deliberately introduced in the data-generating mechanism, our Bayesian model retains reasonable one-step-ahead predictive performance.

\end{document}